\documentclass[aps,prb,twocolumn,showpacs,preprintnumbers,superscriptaddress,longbibliography]{revtex4-2}
\usepackage{graphicx}  
\usepackage{dcolumn}   
\usepackage{bm}        
\usepackage{amssymb}   
\usepackage{framed}
\usepackage{amsmath}
\usepackage{hhline}
\usepackage{enumitem}
\usepackage{subfigure,amsmath,verbatim,moreverb}
\usepackage{tabularx}
\usepackage{lipsum}
\usepackage{longtable}
\usepackage{booktabs}
\usepackage{threeparttable}
\usepackage{booktabs,amssymb,siunitx}
\usepackage{latexsym}
\usepackage{dcolumn}
\usepackage{amsmath}
\usepackage{epsf}
\usepackage{float}
\usepackage{hyperref}
\usepackage{xcolor}
\usepackage[utf8]{inputenc}
\usepackage[normalem]{ulem}
\usepackage{amsmath}
\usepackage{etoolbox}
\usepackage{soul}
\newcounter{subeqn} %

\usepackage{mathptmx}

\usepackage{booktabs}
\usepackage{tabularx}
\usepackage{adjustbox}
\usepackage{rotating}
\usepackage{color,soul}
\usepackage{multirow}
\usepackage{relsize}

\usepackage{hyperref}
\hypersetup{
    unicode=true,          
    pdftoolbar=true,        
    pdfmenubar=true,        
    pdffitwindow=false,     
    pdfstartview={FitH},    
    colorlinks=true,       
    linkcolor=blue,          
    citecolor=blue,        
    urlcolor=blue,
}

\newcommand{\R}{\mathbf{r}}

\usepackage{array}
\newcolumntype{H}{>{\centering\arraybackslash}p{2cm}}

\usepackage[us,12hr]{datetime} 

\begin{document}
\title{Kinetic energy functional constructed from exact gradient expansion of second order in uniform gas limit}
\author{Abhishek Bhattacharjee}
\email{abhishekbhattacharjee179@gmail.com}
\affiliation{School of Physical Sciences, National Institute of Science Education and Research, An OCC of Homi Bhabha National Institute, Jatni 752050, India}
\author{Hemanadhan Myneni}
\email{myneni@hi.is}
\affiliation{Faculty of Industrial Engineering, Mechanical Engineering and Computer Science, University of Iceland, Reykjavík, Iceland}
\author{Manoj K. Harbola}
\affiliation{Department of Physics, Indian Institute of Technology Kanpur, Kanpur 208016, India}

\author{Prasanjit Samal}
\affiliation{School of Physical Sciences, National Institute of Science Education and Research, An OCC of Homi Bhabha National Institute, Jatni 752050, India}

\date{\today}

\begin{abstract}

Orbital-Free Density Functional Theory (OFDFT) has re-emerged as a viable alternative to Kohn–Sham DFT, driven by recent advances in kinetic energy density functionals (KEDFs). Nonlocal (NL) KEDFs have significantly extended OFDFT's applicability, particularly for bulk solids, but their high computational cost and dependence on system-specific parameters limit their universality. In this work, we propose a semilocal KEDF at the Generalised Gradient Approximation (GGA) level that achieves accuracy comparable to the state-of-the-art semilocal functionals, while remaining 
parameter-free. Our construction revives the Thomas–Fermi–von Weizsäcker (TFvW) framework by modulating the relative contributions of TF and vW terms through physically motivated constraints and preserving the exact second-order gradient expansion. Despite its simple form, the proposed functional (KGE2) performs  well across both extended systems (metals and semiconductors) and finite systems (clusters), 
without any need for empirical tuning. 
These results mark a step toward the development of transferable and computationally efficient semilocal KEDFs for large-scale OFDFT simulations. 
\end{abstract}

\maketitle

\section{\label{sec1:Introduction:level1} Introduction }
Since the pioneering work of Hohenberg and Kohn \cite{HK}, later extended by Kohn and Sham \cite{KS}, density functional theory (DFT) has become a cornerstone of \textit{ab initio}  electronic structure calculations. Initially applied to atoms and molecules 
\cite{TF:thomas1927calculation}, DFT 
now underpins research across condensed matter physics, molecular dynamics, warm dense matter (WDM)~\cite{WDM-1-PhysRevLett.113.155006,WDM-2-PhysRevLett.121.145001}, and time-dependent phenomena~\cite{TD-OFDFT-Vignale-causuality,Della-Sala(2022)-TDOFDFT,Pavellano-Kaili-TD-OFDFT_1(2021)_PhysRevB.103.245102,Pavellano-Kaili-TD-OFDFT_2(2021)_PhysRevB.104.235110}.
Its success stems from a compact formalism 
that offers a favorable balance between accuracy and computational efficiency. 

Kohn-Sham (KS) DFT introduces orbitals to compute the kinetic energy exactly, improving accuracy but with a computational cost that scales as $\mathcal{O}(N^3)$ with the system size $N$ \cite{Large-scale-DFT}. In contrast, Orbital-Free DFT (OFDFT) returns to the original Hohenberg-Kohn framework, offering linear $\mathcal{O}(N)$ scaling by solving a single Euler equation \cite{bookdft_Parr-Yang-1989,Review_art_Zhou_2008}:

\begin{align}
    \frac{\delta T_{s}[\rho]}{\delta\rho} + v_{ext}(r) + \int \frac{\rho(\textbf{r}')}{|\textbf{r}-\textbf{r}'|}d^{3}\textbf{r}' + \frac{\delta E_{xc}[\rho]}{\delta\rho} &= \mu
\end{align}

Here, $T_s$ denotes the non-interacting kinetic energy functional, $v_{ext}$ is the external potential, $E_{xc}$ is the exchange-correlation energy functional, and $\mu$ is a Lagrange multiplier fixed from the normalization condition $\int d\textbf{r} \rho(\textbf{r}) = N$ , with $N$ being the total number of electrons. The main challenge in OFDFT is to design kinetic energy density functionals (KEDFs) that are both accurate and transferable across diverse systems.

To address this challenge, significant effort has been devoted to the development of KEDFs, which can be broadly classified into nonlocal and semilocal forms. 
Among these, nonlocal KEDFs (NL-KEDFs) \cite{MGP_2018,KGAP_2017,SM_1993,XMW_PhysRevB.100.205132,WT_1992} 
have achieved considerable success, particularly for systems with inhomogeneous densities \cite{NL-KEDF-PhysRevA.34.2614}.
Most NL-KEDFs retain the von Weizsäcker (vW) term and incorporate linear response based corrections to the Thomas-Fermi (TF) contribution \cite{KGAP_2018}:

\begin{align}
    T^{TF+NL}[\rho] &=  C_{TF} \int\int \rho^{5/6}(\textbf{r})\, \Omega(\textbf{r}-\textbf{r}')\, \rho^{5/6}(\textbf{r}')\, d\textbf{r}\, d\textbf{r}'
    \label{intro_eq2}
\end{align}
where $\Omega(\mathbf{r}-\mathbf{r}')$ is a nonlocal kernel. 
Expanding the kernel about the homogeneous electron gas (HEG) density leads to: 

\begin{align}
    \Omega(\textbf{r}-\textbf{r}') = \delta(\textbf{r}-\textbf{r}') + \omega(\textbf{r}-\textbf{r}') + \cdots
\end{align}
where $\omega(q)$ is typically derived from the HEG linear-response 
function, leading to the general Fourier space form 
\cite{KGAP_2018,Shao(2024)-effective-WT-PhysRevB.110.085129}:

\begin{align}
    \omega(q) = - \frac{\chi_{HEG}^{-1}(\eta) - \chi_{vW}^{-1}(\eta) - \chi_{TF}^{-1}(\eta)}{2\alpha\beta \rho_0^{\alpha+\beta-2}}
    \label{eq:NL-kernel}
\end{align}
with 
$\eta=q/2k_F$ the dimensionless momentum variable, $k_F=(3\pi^2\rho_0)^{1/3}$ is the Fermi wavevector corresponding to the reference density $\rho_0$, and $\chi_{HEG}$, $\chi_{vW}$, and $\chi_{TF}$ are the response functions of the HEG, von Weizsäcker, and Thomas--Fermi models, respectively. The parameters 
$\alpha, \beta$ are real numbers. 
Although NL-KEDFs   
have achieved high accuracy for many bulk systems, particularly metals, their transferability to 
semiconductors and finite systems can depend strongly on the kernel construction and the treatment of the reference density~\cite{KGAP_2018,review-article-OFDFT-LargeScale}.  
Moreover, some NL-KEDFs involve system-dependent parameters or reference-density choices that can limit transferability~\cite{JGM-10.1063/5.0204957}.  
As Kohn noted, semilocal functionals can effectively capture nonlocal effects when only short-range interactions are relevant, a concept referred to as 'shortsightedness' \cite{nearsightedness_Kohn_2005}.

Alternatively, the development of improved semilocal functionals that can mimic the accuracy of NL  approaches remains an active area of research. 
Despite its simplicity, within the BLPS pseudopotential framework, the Thomas–Fermi–von Weizsäcker (TFvW) functional has been reported to  predict bonding  in  simple metals and semiconductors~\cite{PGSL_2018_doi:10.1021/acs.jpclett.8b01926}, whereas several NL-KEDFs, such as the Wang-Teter (WT)~\cite{WT_1992} and Wang-Govind-Carter (WGC)~\cite{WGC_1998,WGC_1999} forms, fail to do so for some of these systems~\cite{MGP_2018,Elastic_carter,LKT-SBTrichey-GGA_PhysRevB.98.041111}. This bonding behavior should be understood as arising within the combined KEDF/BLPS framework rather than from the TFvW approximation alone.

The general form of the TFvW functional is 
\begin{align}
    T_s[\rho] = T_{TF}[\rho] + \lambda T_{vW}[\rho]
\end{align}
with $\lambda \in [0,1]$. The vW term ensures correct exponential decay, satisfies the Kato cusp conditions, and captures the short-wavelength linear response behavior \cite{KATO_cusp1,bookdft_Parr-Yang-1989,KATO_cusp2,JonesAndGunnarsson,Jones2015}.

The construction of semilocal KEDFs is further motivated 
by the conjointness conjecture \cite{Conjointness-conj-PhysRevA.44.768}, which links the forms of exchange and kinetic enhancement factors. According to this conjecture, any exchange density functional approximation  (DFA) can, in principle, be transformed into a KEDF \cite{APBEk,APBE}. 
However, not all such functionals satisfy exact constraints ~\cite{AugPC_semilocal_KEDF_molecoule_2024}. A comprehensive list of GGA-type semilocal KEDFs is given in Ref.~\cite{AugPC_semilocal_KEDF_molecoule_2024}. Most of these functionals fail to improve local quantities such as the electron density 
and often rely on error cancellation for accurate total energy predictions \cite{Pearson_is_good_10.1063/1.2774974,AugPC_semilocal_KEDF_molecoule_2024}.


A notable example illustrating these limitations is the APBEK functional \cite{APBEk}, inspired by the success of the PBE exchange functional. However, it fails to reproduce the correct second-order gradient expansion (GE2) coefficient and does not satisfy Pauli positivity. 
Its refinement,  VT84 \cite{VT84_PhysRevB.88.161108} by Karasiev et al., addressed several of these 
shortcomings and marks the beginning of competitive semilocal KEDFs. 
Since then, a variety of functionals have been proposed, 
some optimized for metals (Luo-Karasiev-Trickey  (LKT) \cite{LKT-SBTrichey-GGA_PhysRevB.98.041111,LKTF-SBTrichey-GGA_PhysRevB.101.075116}), and others tailored for semiconductors or localised systems, such as the Pauli–Gaussian (PG) family
\cite{PGSL_2018_doi:10.1021/acs.jpclett.8b01926,PGSL_2019_assessment,GSE2_2019_Lucian_PhysRevB.99.155137,GSE2_2019_Lucian_PhysRevB.99.155137}).
Among these, the PG1 functional is one of the most accurate semilocal KEDFs. However, it does not recover the correct gradient expansion in the low-$s$ limit. 
Restoring the exact GE2 coefficient 
by setting $\mu=40/27 \approx 1.481$ improves formal consistency but deteriorates the predicted mechanical properties, particularly for metals.
PGS denotes this GE2-recovering
variant of the PG functional. Further extensions, such as PGSL and PGSLr \cite{PGSL_2019_assessment}, incorporate meta-GGA (mGGA) ingredients to  improve bulk properties and broaden applicability. 
These improvements come at the cost of increased computational cost and numerical instabilities, especially in finite systems. 
Such issues are particularly pronounced for clusters, where periodic implementations may  introduce noise in Laplacian-dependent terms. 

As discussed above, the NL-kernels often rely on the Fermi momentum $k_F$ (see Eq. \ref{eq:NL-kernel}), typically approximated as a 
constant, $k_F = (3\pi^2\rho_0)^{1/3}$, where $\rho_0$ is the average electron density, to reduce computational cost. Only a few density-dependent kernels, such as HC \cite{HC}, revHC~\cite{revHC_2021}, LMGP~\cite{LMGP-00} and LJGM~\cite{LJGM_JCTC_Bhattacharjee_2026}, employ a local description, $k_F = (3\pi^2\rho(\mathbf{\R}))^{1/3}$, which 
increases the computational expense. 
However, the use of a fixed reference density in WT-class density-independent NL kernels can become problematic for finite systems~\cite{review-article-OFDFT-LargeScale,Review_article_2024_Ma_Mi_Wang,LMGP-00,LJGM_JCTC_Bhattacharjee_2026}, where the choice of an average density may depend on the simulation volume.
Nevertheless, previous molecular-dynamics studies of sodium clusters using density-independent kernels have reported promising results~\cite{Na-cluster-Aguado-2005-PhysRevLett.94.233401,Na-melting-Aguado-2006-PhysRevB.74.115403}.

Motivated by the need for a simple, transferable, and computationally efficient kinetic energy functional, we revisit semilocal KEDFs and  
propose a parameter-free functional designed to provide balanced accuracy across different classes of systems. 
We benchmark our proposed GGA-level functional, KGE2, against state-of-the-art NL- and mGGA-KEDFs. 
KGE2 achieves 
competitive accuracy without the use of  fitting parameters and demonstrates good transferability across both bulk and finite systems.

The rest of the article is organized as follows: Section~\ref{sec:Theory} describes  the construction of KGE2. Section~\ref{Sec:Result:level-1} discusses benchmark results for bulk systems. Section~\ref{sec:Result:level-5-clusters} presents an assessment for molecular clusters. Finally, conclusions are presented in Section~\ref{sec5-conclusion}.


\section{\label{sec:Theory} Theory}
\subsection{\label{sec:Theory:level-1} Functional construction }
 
Following the constraint-based approach of Perdew,\cite{PERDEW1992} we seek to construct a simple, nonempirical GGA-level KEDF that satisfies several known exact constraints and exhibits broad applicability across both bulk and finite systems.
To this end, we introduce the reduced density gradient $s=|\nabla\rho(r)|/2k_{F}\rho(r)$ as a key dimensionless variable. The generic GGA form of Eq.\ref{intro_eq2} can thus be written in terms of $s$ as:
\begin{align}
    T_{s}^{GGA} &= \int ~\tau_{TF}(\rho)F_{\theta}(\rho,s) d\textbf{r} + \lambda~ T_{vW}  \nonumber \\
    & = \int  \underbrace{ ~\tau_{TF} [F_{\theta}(\rho,s) + \lambda\frac{5}{3}s^{2} ] }_{\tau_{s}} d\textbf{r} 
    \label{Eq:ggaTs}
\end{align}
where $\tau_{TF}=C_{TF}\rho^{5/3}$ is the Thomas Fermi kinetic energy density and  $C_{TF}=\frac{3}{10}(3\pi^{2})^{2/3}$. Here, $F_{\theta}$ denotes 
the Pauli enhancement factor, emphasizing that the vW term does not incorporate any information related to the Pauli exclusion principle (PEP)~\cite{bookdft_Parr-Yang-1989,review-article-OFDFT-LargeScale}. Throughout this work, we set $\lambda=1$.

\begin{figure}[h!]
    \centering
    \includegraphics[scale=0.3]{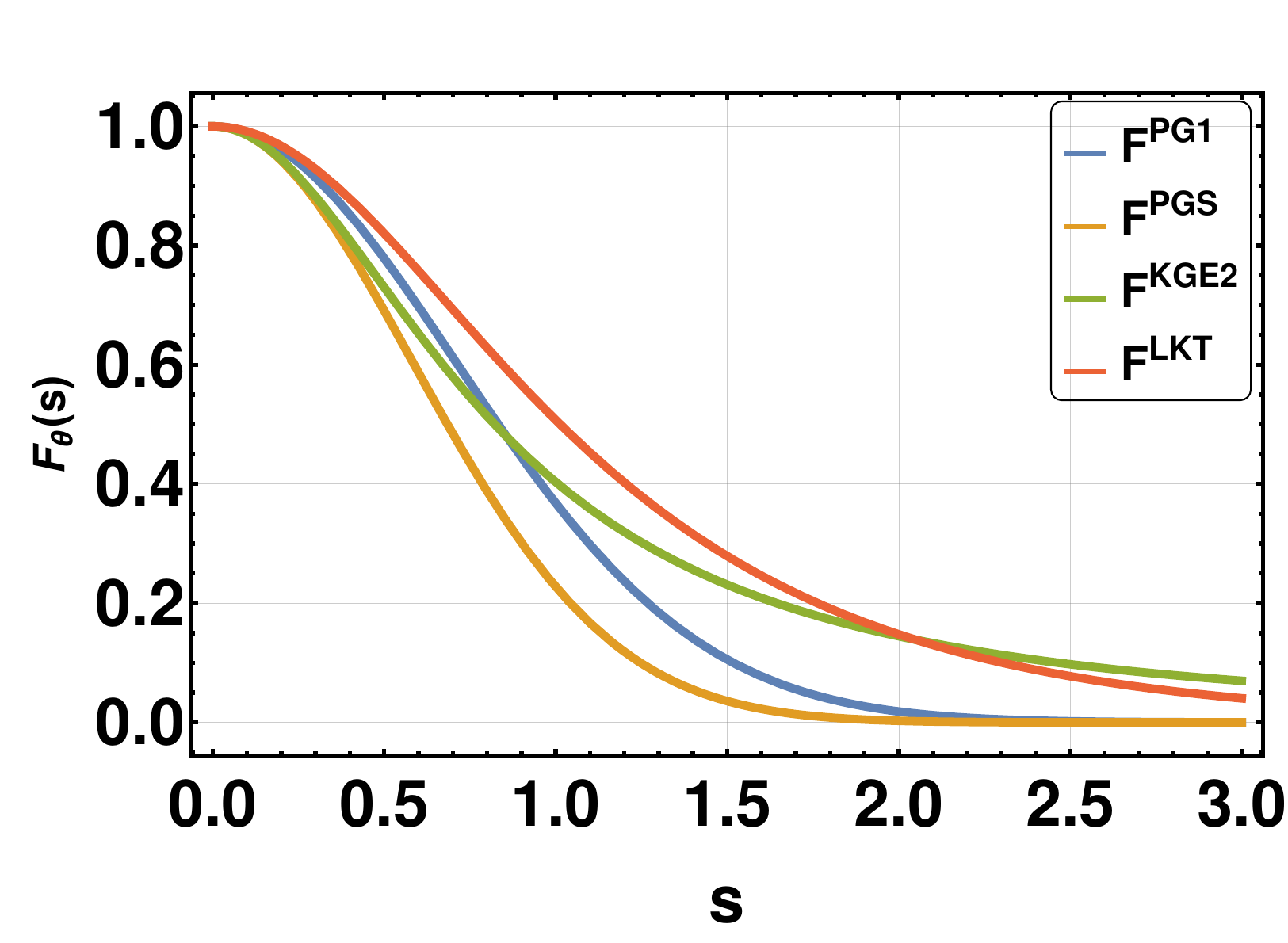}
    \caption{Plot of Pauli enhancement factors of various functionals as a function of variable $s$.}
    \label{fig:F_theta_enhance}
\end{figure}

The construction of the functional must satisfy the correct asymptotic behavior, guided by the following conditions:
\begin{itemize}
    \item $\lim_{s\to\infty} \tau_{s} \rightarrow \tau_{vW} $ 
    \item $\lim_{s\to 0} \tau_{s} \rightarrow \tau_{TF}+\frac{1}{9}\tau_{vW} $ 
    \item $\tau_{s} \ge 0 $ for all \textbf{r} 
\end{itemize}
where $\tau_{vW} = \frac{|\nabla \rho|^2}{8\,\rho} $ is the von Weizsäcker kinetic energy density.
The second condition corresponds to recovery of the second-order gradient expansion (GE2) in the slowly varying density limit. Motivated by these constraints, 
we propose a new Pauli enhancement factor, hereafter referred to as KGE2, 
\begin{align}
    F_{\theta}^{KGE2} = \frac{1}{1+\alpha s^{2}} 
    \label{eq:F_tr02}
\end{align} 
where the parameter $\alpha$ is chosen as 1.481 to recover the  second-order gradient expansion approximation (GE2) in the small-$s$ limit. As mentioned in Sec.~\ref{sec1:Introduction:level1},  previously proposed similar semilocal enhancement factors, namely PG and LKT, have the forms $F_{\theta}^{PG}=\exp{(-\mu s^2)}$ ~\cite{PGSL_2018_doi:10.1021/acs.jpclett.8b01926} and $F_{\theta}^{LKT}=\frac{1}{\cosh(a s)}$ ~\cite{LKT-SBTrichey-GGA_PhysRevB.98.041111}, respectively. 
For comparison, Fig.~\ref{fig:F_theta_enhance} shows the Pauli enhancement factors of
PG1, PGS, KGE2, and LKT. The low-$s$ expansion shows that PGS
and KGE2 recover the GE2 limit, thereby satisfying the second
constraint,
$\lim_{s\to0}\tau_s=\tau_{\rm TF}+\frac{1}{9}\tau_{\rm vW}$.
A key feature of KGE2 is its exact recovery of the GE2 behavior
in the low-$s$ regime while retaining a simple parameter-free
rational form.

The resulting new kinetic energy density functional 
is given by 
\begin{align}
    \label{eq:try03}
    \tau_{s}^{KGE2} = 
    \tau_{TF}[ F^{KGE2}_{\theta}+\frac{5}{3}s^{2} ]
    = \tau_{TF} & \frac{1}{1+\alpha s^{2}} + \tau_{vW} 
\end{align}

The corresponding Pauli potential is defined as
\begin{align}
    v_{\theta}(\textbf{r}) &= \frac{\delta }{\delta\rho(r)}( T_{s}[\rho] - T_{vW}[\rho] ) \nonumber 
\end{align}
while the full kinetic potential becomes 
\begin{align}
    v_{s} &= v_{\theta} + v_{vW}, \nonumber 
\end{align}
with 
\begin{align}
v_{vW}(\textbf{r}) &= \frac{1}{8} \Big( \frac{(\nabla\rho(\textbf{r}))^{2}}{\rho^{2}(\textbf{r})} - 2\frac{\nabla^{2}\rho(\textbf{r})}{\rho(\textbf{r})} \Big) \nonumber
\end{align}
The explicit form of $v_{\theta}$ for a generic semilocal class of $T_{s}^{GGA}$ as in Eq.\ref{Eq:ggaTs} is given by:
\begin{align}
    v_{\theta} &= \frac{\partial}{\partial\rho} (\tau_{TF}F_{\theta}) - \nabla \frac{\partial}{\partial(\nabla\rho)} (\tau_{TF}F_{\theta} ) 
\end{align}
Substituting the KGE2 enhancement factor of Eq.~\ref{eq:F_tr02} yields    
\begin{align}
v^{KGE2}_{\theta}(\textbf{r}) = &\frac{1}{3} \frac{C_{TF} \rho^{2/3}(\textbf{r})}{(1+\alpha s^{2})^{2}} [ 5+11\alpha s^{2} ]  + \frac{2}{3} \frac{\alpha C_{TF}\rho^{2/3}(\textbf{r})}{(1+\alpha s^{2})^{3}}  \nonumber \\
     & ~ \times\Big[ -3s^{2} + 13\alpha s^{4} + \frac{3\nabla^{2}\rho(\textbf{r})}{c_s^{2}\rho^{5/3}(\textbf{r})} \{1-9\alpha s^{2} \}  \Big] 
     \label{eq:v_theta_try02}
\end{align}

The resulting Pauli potential satisfies several known exact constraints associated with the Pauli potential, as discussed in Ref.~\cite{Exact-pauli-Levy_PhysRevA.38.625} and elaborated in Section~\ref{sec:Theory:level-3-pauli}. 

\subsection{\label{sec:Theory:level-2-LinearResponse} Response function analysis of semilocals }

One of the key features that distinguishes NL-KEDFs from semilocal-KEDFs is their  ability to reproduce 
the linear response behavior of the HEG~\cite{review-article-OFDFT-LargeScale,Hemanadhan_2012}. 
In contrast, semilocal KEDFs can recover the correct long-wavelength (small-$q$) limit only when the exact small-$s$ behavior is enforced. 
The derivation of the linear response for semilocal KEDFs is more intricate than that for NL-KEDFs since the conventional definition of response function, $\frac{1}{\chi(\R-\R')} = \frac{\delta^2 T[\rho]}{\delta\rho(\R)\delta\rho(\R')}$, is not directly applicable. 
To analyze the response, we consider a small perturbation in the uniform electron gas (UEG) with  density
$\rho=\rho_0 + \rho_q e^{i \mathbf{q}\cdot \mathbf{r}}$, where $\rho_q/\rho_0 << 1$. At $\mathbf{r}=0$, 
$\rho=\rho_0+\rho_q$, $\nabla\rho=i\mathbf{q}\rho_q, ~ \nabla^2\rho=-q^2\rho_q$. 
Starting from the Pauli potential expression in Eq.\ref{eq:v_theta_try02}, and following the derivation of the response function given in Appendix.\ref{App:Linear_Response}, 
the inverse response function for KGE2 becomes 
\begin{align}
    \frac{1}{\chi^{KGE2}} & = \frac{\pi^{2}}{k_{F}} - \frac{3}{5}\alpha\eta^2 \frac{\rho_q}{\rho_0} + \frac{18}{5} \frac{\pi^2}{k_F}\alpha \eta^2 (\frac{\rho_q}{\rho_0}) \Big(1 - 8\frac{\rho_q}{\rho_0} \Big) \nonumber\\ &~~~~~~~~~~~~~  - \frac{9}{5} \frac{\pi^2 }{k_F} \eta^2 \Big( 1 + 3\frac{\alpha q^2}{k_{F}^2} (\frac{\rho_q}{\rho_0})^2 \Big) \nonumber 
\end{align}    
Retaining only terms independent of $\rho_q$ yields the linear-response limit, 
\begin{align}
    \frac{1}{\chi^{KGE2}} &= \frac{\pi^{2}}{k_{F}} \big[ 1 - \frac{9}{5}\alpha\eta^2 + \mathcal{O}(\rho_{q}) \big] 
    \label{eq:Chi_full_GGA}
\end{align}
where $\eta= \frac{q}{2k_F}$. Including the vW contribution, $\frac{1}{\chi^{vW}}=\frac{\pi^2}{k_F}3\eta^2$, the general response of a GGA-class semilocal KEDF becomes
\begin{align}
    \frac{1}{\chi_{GGA} } &= -\frac{\pi^2}{k_{F}}(\frac{9}{5}\alpha\eta^2 - 3\eta^2 - 1) \\
    \frac{1}{\chi_{\theta}^{GGA}} &= \frac{1}{\chi_{GGA}} - \frac{1}{\chi^{vW}}- \frac{1}{\chi^{TF}} \\
    (\chi_{\theta}^{GGA})^{-1}&= -\frac{\pi^2}{k_F} \frac{9}{5}\alpha\eta^2 \\
    G^{GGA}_{\theta}(\eta) &= \frac{\chi^{TF}}{\chi^{GGA}_{\theta}} = - \frac{9}{5}\alpha\eta^2
    \label{eq:Chi_GGA}
\end{align}

\begin{figure}
    \centering
    \includegraphics[width=1.0\linewidth]{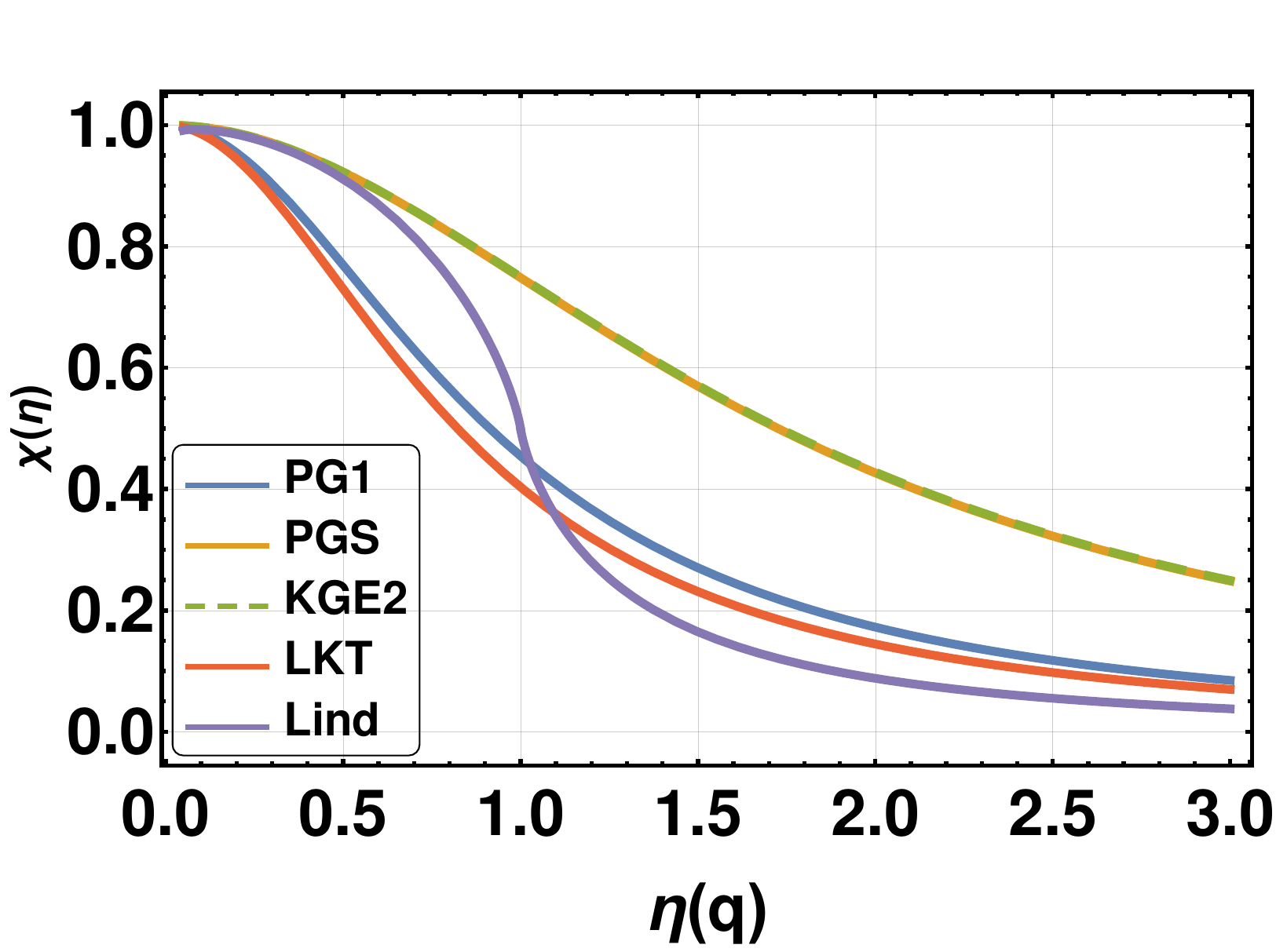}
    \caption{Comparison of the density-response functions of various semilocal KEDFs with the exact Lindhard response function.}
    \label{fig:Chi_GGA_vs_Lind}
\end{figure}


The ability of a semilocal KEDF 
to reproduce the Lindhard response depends critically on the form of the enhancement factor and, in particular, on recovery of the exact GE2 coefficient. 
Notably, several GGA and meta-GGA KEDFs, including PGSL and PGSLr, exhibit the same quadratic small-\(\eta\) response behavior as Eq.~\ref{eq:Chi_GGA}, differing only in the prefactor \(\alpha\)~\cite{PGSL_2019_assessment}.

In the limit $s\to0$, corresponding to the long-wavelength limit $q\to0 $ or $\eta\to0$, both PGS and KGE2 recover the exact GE2 expansion and therefore reproduce the correct small-wavevector limit of the Lindhard response.
Since the derivation of GGA response is based on the slowly varying density limit  $(s\to0)$, any meaningful comparison with the exact Lindhard response is only valid near the vicinity of $\eta\to0$. Beyond this regime, the semilocal response curves shown in Fig.~\ref{fig:Chi_GGA_vs_Lind} represent extrapolations of the small-$\eta$ behavior rather than 
exact representations of the full HEG response. Importantly, no semilocal GGA-type KEDF can reproduce inherently nonlocal features of the exact Lindhard response, such as the logarithmic singularity near \(\eta=1\), which gives rise to Friedel oscillations~\cite{GSE2_2019_Lucian_PhysRevB.99.155137}. 
The results presented in Fig.~\ref{fig:Chi_GGA_vs_Lind} demonstrate that enforcing the exact GE2 coefficient leads to improved  agreement with the Lindhard response in the small-\(\eta\) regime, whereas empirical deviations from the GE2 constraint introduces larger deviations. Similar trends are also observed in the density analysis discussed in Sec.~\ref{sec:Result:level-4-Density-analysis}.


\subsection{\label{sec:Theory:level-3-pauli} Pauli potential and exact constraints }
 
The Pauli potential plays a central role in OFDFT since it is solely responsible for the effects of Pauli-exclusion principle on kinetic part of the energy. In many ways the Pauli potential is similar to XC potential in KS-DFT, the exact expression of the Pauli potential is not known and hence needed to be approximated. Also, the quality of the electron density is closely related to the behavior of the Pauli potential~\cite{Pauli-Density-connection_10.1063/1.4940035}, making it an important diagnostic quantity in OFDFT.

Early work by Levy and Ou-Yang ~\cite{Exact-pauli-Levy_PhysRevA.38.625} established several exact conditions and scaling relations associated with the Pauli potential.
These constraints have subsequently served as important guidelines in the construction of semilocal kinetic-energy density functionals (KEDFs). Related scaling-based approaches were later explored by Nagy~\cite{Nagy-PauliPot-PhysRevA.84.032506,scaling4,Atomic_Pauli_pot_2021}.

The Pauli enhancement factor employed in semi-local functionals highlights that the von Weizsäcker (vW) term alone does not account for the effects of the Pauli exclusion principle (PEP). As a result, the atomic or solid-state model described solely by the vW functional corresponds to a single-orbital system, with the effects of the PEP introduced as a corrective term. This correction is introduced via  the Pauli potential, denoted by $v_{\theta}$. 
The 
$T_{vW}$ provides a  lower bound to the non-interacting kinetic energy: $T_{s} \ge T_{vW}$. Hence,  incorporating the effects of PEP through the Pauli kinetic energy contribution $T_{\theta}$ will increase the kinetic energy, leading to the important constraint: $T_{\theta} \ge 0 $. In 2011, Nagy \cite{Nagy-PauliPot-PhysRevA.84.032506, scaling4,Atomic_Pauli_pot_2021} proposed a method for constructing an approximate Pauli potential based on density scaling relations. 
More generally, one of the standard strategies for constructing approximate functionals is the enforcement of exact or physically motivated constraints. 
Here, we summarize some of the most important constraints~\cite{Exact-pauli-Levy_PhysRevA.38.625}. 
\begin{enumerate}[label=(\alph*)]
    \item $F_{\theta} \ge 0  $
    \item $v_{\theta}(\textbf{r}) \ge 0 $ for all \textbf{r}
    \item $t_{\theta}(\R) \ge 0 $ for all \textbf{r}
    \item $v_{\theta}(\textbf{r}) \ge t_{\theta}(\R) $ for all \textbf{r} $\qquad \qquad $
    \item $\frac{ \tau_{\theta}[\rho_\lambda]}{\rho_{\lambda}} = \lambda^{2}~ \frac{\tau_{\theta}[\rho_\lambda]}{\rho(\lambda\textbf{r})} \qquad \qquad$
    \item $v_{\theta}[\rho_{\lambda}](\textbf{r}=\textbf{r}_{0}) = \lambda^{2}v_{\theta}[\rho](\textbf{r}=\lambda \textbf{r}_{0}) $
\end{enumerate}

Here, $t_{\theta}=t_{s}-t_{vw}=\frac{\tau_\theta}{\rho} $ denotes the Pauli-kinetic-energy-density, where $T_{s}=\int t_{s}(\R)~n(\R)~d\R$. And, 
$\lambda\in[0,1] $ is the uniform coordinate scaling parameter, which provides insight into how the functionals behave as the system transitions between high- and low-density regimes.

For the present KGE2 functional, the positivity of \(F_{\theta}\) and \(v_{\theta}\) follows directly from Eq.~\ref{eq:F_tr02} and Eq.~\ref{eq:v_theta_KGE2}, respectively. Furthermore, because the enhancement factor depends only on even powers of the reduced density gradient \(s\), the Pauli kinetic-energy density, $t_{\theta}$, remains positive within the physically relevant density regime. In the low-\(s\) limit, KGE2 recovers the second-order gradient expansion (GE2), thereby satisfying the corresponding uniform coordinate scaling relations.
Under uniform coordinate scaling, the reduced density gradient transforms as 
$s[\rho_{\lambda}](\mathbf r) = s[\rho](\lambda \mathbf r).$
Consequently, 
\begin{align}
    \frac{\tau_{\theta}^{KGE2}}{\rho}[\rho_{\lambda}] &= \lambda^{2} C_{TF} \frac{\rho^{2/3}(\lambda(\textbf{r}))}{1+\alpha s^{2}(\lambda(\mathbf{r}))} \nonumber \\
    \frac{\tau_{\theta}^{KGE2}}{\rho}[\rho_{\lambda}](\mathbf{r}=\mathbf{r}_{0}) &= \lambda^2\frac{\tau_{\theta}^{KGE2}}{\rho}[\rho](\mathbf{r}=\lambda \mathbf{r}_{0})
\end{align}
Similarly, it can be shown that 
\begin{align}
    v_{\theta}^{KGE2}[\rho_{\lambda}](\mathbf{r}=\mathbf{r}_{0}) = \lambda^{2} v_{\theta}^{KGE2}[\rho](\mathbf{r}=\lambda \mathbf{r}_{0})
\end{align}
Further details are provided in Appendix.\ref{App:Exact-Scaling}.
In addition to these exact scaling conditions, one may also examine the Lieb–Simon scaling behavior~\cite{Lieb-Simon-Scaling-1973,Atomic_Pauli_pot_2021}, which provides a significantly more stringent criterion for approximate kinetic energy functionals.

%

\section{Results and discussions\label{Sec:Result:level-1}}

\subsection{ Computational Details and Materials \label{sec:Result:level-2-comp.details}}

All the OFDFT calculations were performed 
using the DFTpy software package~\cite{DFTpy:Shao_2020}. 
Throughout this work, we have used the LDA level bulk-derived local pseudopotential (BLPS)~\cite{BLPS-HC, DFTpy:Shao_2020} of Huang and Carter.
Previous assessments of semilocal KEDFs have shown that the qualitative trends of structural and energetic properties in solids are relatively insensitive to the specific choice of pseudopotential~\cite{PGSL_2019_assessment}. 
In the BLPS pseudopotential framework employed throughout this work, 
core electrons are frozen and valence densities remain  smooth. The BLPS defines the valence electron-ion interaction, while the KEDF provides the approximate noninteracting kinetic energy and the associated Pauli potential within this valence-space framework. We have used the LSDA exchange-correlation~\cite{PW91} in all our calculations. 
For consistency across all calculations, we adopted the 
universal parameter for HC ($\lambda=0.01177, \alpha=1.952$ and  $\beta=0.7143$) from Ref.~\cite{revHC_2021}, $a=1.3$ for the LKT functional~\cite{LKT-SBTrichey-GGA_PhysRevB.98.041111}, $\mu=1.0$ for PG1, and $\mu=40/27$ for PGS \cite{PGSL_2019_assessment}. 
In this work, all OFDFT calculations were performed using a kinetic energy cutoff of $1200$ eV.

\textbf{Solids:} 
For the benchmark calculations of metals, we considered simple metals: Al, Li, and Mg, each with body-centred cubic (BCC), face-centred cubic (FCC), and simple-cubic (SC) phases. For semiconductors, we examined nine III-V cubic zincblende (ZB) semiconductors. 
We compute the equilibrium volume ($V_0$), bulk moduli (B$_0$), and equilibrium total energies of all those aforementioned systems. 
We use the Birch-Murnaghan first-order equation of state~\cite{BrichMurnaghan} to fit the energy versus volume curve. Reference KS results for solids were taken from available literature \cite{HC,KGAP_2018}. 

\textbf{Clusters:} 
For calculation of cluster systems, we have considered: Mg$_8$, Mg$_{30}$, Mg$_{50}$, Li$_8$, Li$_{30}$, Al$_8$, Si$_8$, Si$_{30}$,  Al$_4$P$_4$, Al$_4$Sb$_4$, Ga$_4$As$_4$, Ga$_{10}$As$_{10}$, Ga$_4$P$_4$, Ga$_4$Sb$_4$, In$_4$As$_4$, In$_4$P$_4$ and In$_4$Sb$_4$.  
Some of the structures are provided in the supporting information \cite{support}.  
All the rest of the cluster structures are generated by CALYPSO-5.0 code \cite{CALYPSO-1-WANG20122063,CALYPSO-2-PhysRevB.82.094116,CALYPSO-3-10.1063/1.4746757} with VASP \cite{vasp1} interface. Each isolated cluster was placed in an orthorhombic simulation box with periodic boundary condition. 
The separation between the cluster and its nearest neighboring periodic images was set to 16 \AA, which is sufficiently large to ensure that interactions 
neighboring images are negligible. 
The most stable structure was selected, and both KS-DFT and OFDFT calculations were performed on it.
%
During structure generation, the minimal and maximum interatomic distances  
were constrained within approximately 0.7 \AA to 2.8 \AA, depending on the system. 
%
%
Benchmark KS-DFT calculations were carried out using Quantum-ESPRESSO (QE) \cite{QuantumEspresso-Giannozzi_2009}.
A plane-wave cutoff of 30 Ry was sufficient to converge the total energies to within approximately 1 meV/atom using the same simulation cells employed in the OFDFT calculations.

To compare electron densities obtained from different KEDFs and KS-DFT, all  densities were interpolated onto uniform (100, 100, 100) real-space mesh grids for analysis. 


\subsection{\label{sec:Result:level-4-Density-analysis} Analysis of electron density}

\begin{figure}[h!]
    \centering
    \includegraphics[scale=0.35]{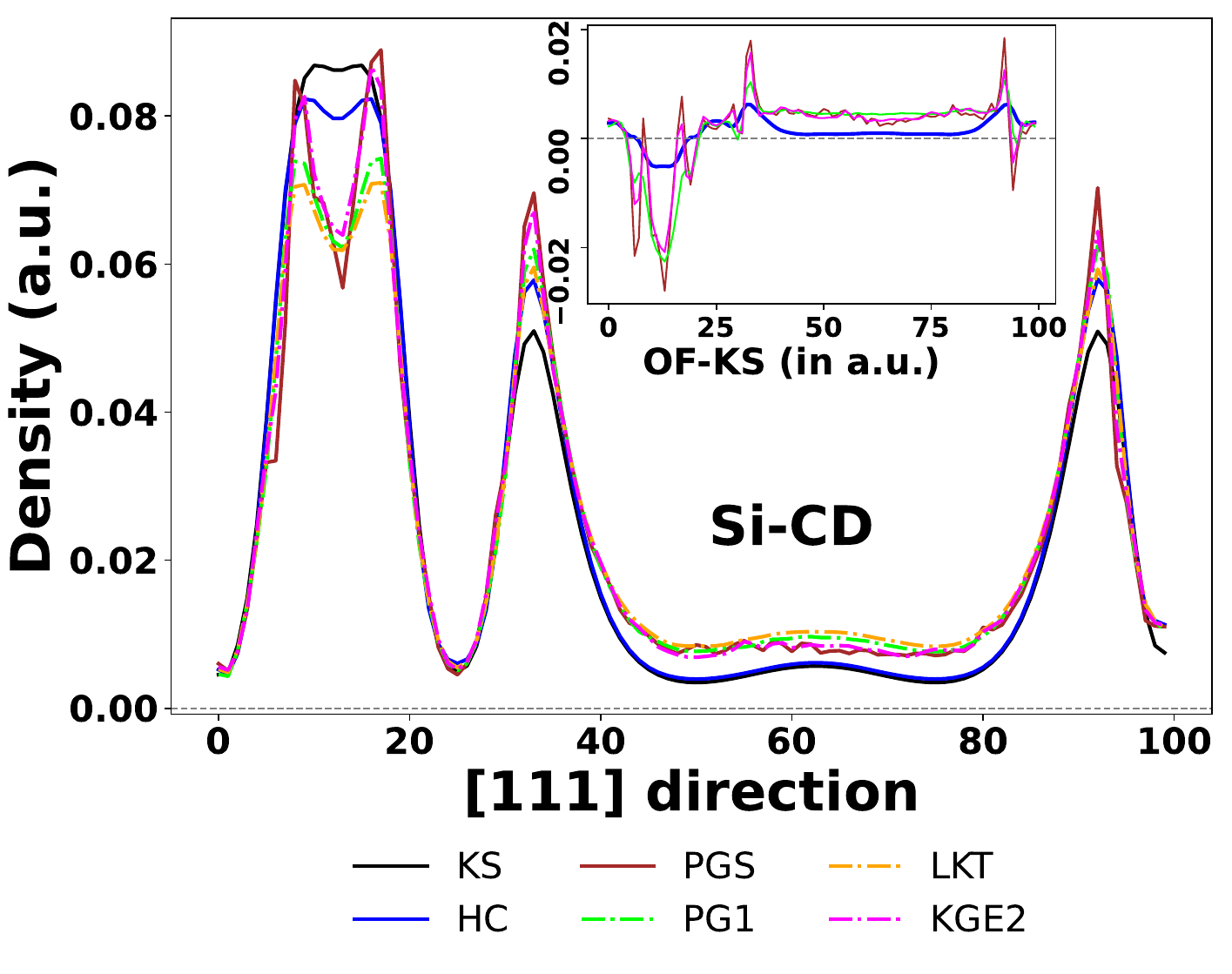}
    \caption{Density profile of the Si CD along [111] direction for various available 
    KEDFs. The blue line is KS-density; see legend for other labels. The inset plot is density difference $\rho^{KS}-\rho^{OF}$ along [111] direction. The density is in atomic units, and on the x-axis, we have grid points. }
    \label{fig:DenPlot_CD}
\end{figure}

The accuracy of the electron density is as important as, if not more important than, the prediction of total energies. A reliable kinetic energy functional should improve both simultaneously \cite{review-article-OFDFT-LargeScale}. It is well known that one of the advantages of nonlocal (NL) KEDFs is their improved accuracy in predicting the density compared to semilocal functionals. As shown in Fig.~\ref{fig:DenPlot_CD}, both KGE2 and PGS yield significantly improved densities relative to the TFvW. 

The interstitial region is particularly important  because it corresponds to the bonding region between atoms. This is also where long-range effects or low-momentum ($q \to 0$) become relevant, the vW term vanishes due to flat density, and the accuracy of the enhancement factor becomes especially important. 
Among the NL-KEDFs designed to recover the correct low-$q$ behavior, the jellium-with-gap (JGM) functional~\cite{JGM-10.1063/5.0204957} and the Huang and Carter (HC) functional~\cite{HC} are notable examples. Comparing our results with Kohn–Sham (KS) and HC densities, we find that semilocal KEDFs generally tend to overestimate the density in regions far from the atomic centers. 

In the original work of the Pauli-Gaussian (PG) functional~\cite{PGSL_2018_doi:10.1021/acs.jpclett.8b01926}, the optimized value $\mu = 1$ was reported to provide the best overall performance. However, 
our density analysis suggests that 
$\mu = 40/27$, corresponding to recovery of the exact second-order gradient expansion (GE2), results in better agreement with KS densities. 
More generally, 
semilocal KEDFs that recovers the exact GE2 limit tend 
to provide more accurate density than versions 
to provide obtained through empirical parameter tuning.
Even the addition of meta-GGA corrections, as in PGSL and PGSLr~\cite{PGSL_2019_assessment}, degrades density quality in the interstitial regions 
relative to PGS and KGE2 (see Fig.~2 in Ref.~\cite{support}). This observation suggests that the inclusion of higher-order terms can worsen the local density description,  due to parameter choices optimized for either metals or semiconductors, but not both.

\begin{figure*}[htbp]
    \centering
    \includegraphics[scale=0.9]{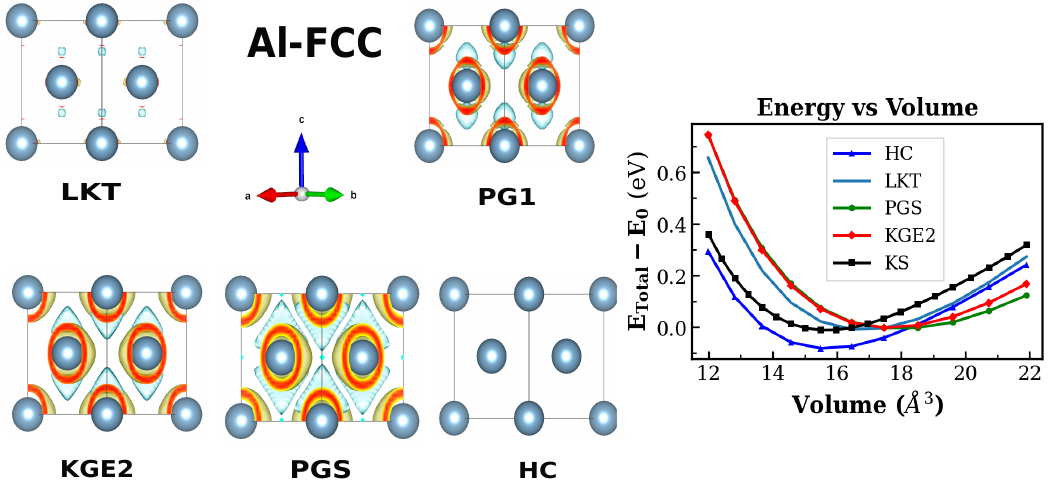}
    \caption{Density difference (left) of Al-FCC (KS-OF) in 3D in the order: 
    LKT, PG1, KGE2, PGS, and HC.
    Isoscale is set at 100\% yellow saturation for density difference of 0.02 $a.u$ and 100\% blue saturation for density difference -0.02 $a.u.$. On the right side we have plot of total energy as function of changing volume for various KEDFs. Here we have shifted the energy values with KS equilibrium energies ($E_0$) for meaningful comparison.  
    }
    \label{fig:3d-dendiff-Al-FCC}
\end{figure*}

\begin{figure*}[htbp]
    \centering
    \includegraphics[scale=0.9]{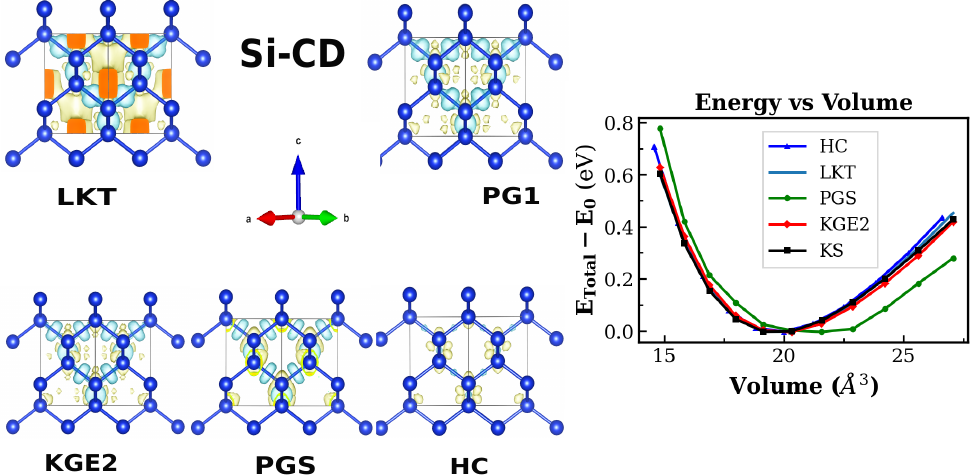}
    \caption{Density difference of Si-CD (KS-OF) in 3D in the order: 
    LKT, PG1, KGE2, PGS, and HC. 
    Isoscale is set at 100\% yellow saturation for density difference of 0.035 $a.u$ and 100\% blue saturation for density difference -0.035 $a.u.$. On the right side we have plot of energy as function of changing volume for various KEDFs. Here we have shifted the energy values with KS equilibrium energies ($E_0$) for meaningful comparison.
    }
    \label{fig:3d-dendiff-Si8}
\end{figure*}

For metallic systems, LKT yields the smallest density errors, followed by KGE2 and then PG1, as shown in Fig.~\ref{fig:3d-dendiff-Al-FCC}. For comparison, 
the corresponding 
three-dimensional density difference for crystalline silicon are presented in Fig.~\ref{fig:3d-dendiff-Si8}, 
revealing similar trends and reinforcing the above conclusions.

The consistent overestimation of density in interstitial regions suggests a fundamental limitation of the semilocal class of functionals. These regions are characterized by low and relatively uniform densities, where the vW 
contribution becomes small 
and the Pauli term dominates. 
This behavior suggests 
that the Pauli potential in GGA-level KEDFs may be underestimated in these areas likely due to a TF component that decays too rapidly (see Fig.~4 in Ref.~\cite{support}) in small-$s$ region.


A more stringent measure of density quality is the density error metric $D_0$ \cite{PGSL_2018_doi:10.1021/acs.jpclett.8b01926}, defined as:
\begin{align}
    D_0 = \frac{1}{N_e} \int | \rho^{\mathrm{OFDFT}}(\textbf{r}) - \rho^{\mathrm{KSDFT}}(\textbf{r}) | \, d^3\textbf{r},
\end{align}
where $N_e$ is the total number of electrons per unit cell. 
This metric quantifies the deviation of the OFDFT density from the KSDFT reference density.
To be consistent with the other relative errors, the reported $D_0$ values are expressed as percentages (\%) by multiplying by 100.
 The $D_0$ values for all systems are provided in Table IV of the Supporting Information~\cite{support} and are summarized in Fig.~\ref{fig:box_all}. 
Representative values for selected systems are also presented in Table~\ref{tab:Bulk-data}.

In general, density errors for semiconductors are 
substantially larger than those for metals. Among the semilocal functionals considered here, PGS has the lowest mean $D_0$ error for semiconductors (16.23\%), followed closely by KGE2 (16.93\%) and LKT (18.63\%) (see Fig.~S1 of the Supporting Information~\cite{support}). For comparison, HC achieves a substantially lower mean $D_0$ error of 11.48\%. In metals, LKT performs best (2.95\%), followed by KGE2 (4.75\%) and PGS (7.15\%). 
Overall, for the combined semiconductors and metals dataset, the density errors of 
KGE2 (14.5\%) lies between PGS (13.8\%) and LKT (15.1\%) as can be seen from Fig.~\ref{fig:box_all}. 

Although the mGGA functionals, particularly PGSLr, improve density predictions for metals, their performance is comparable to LKT 
and is consistent with earlier studies (Table 3 of Ref.~\cite{PGSL_2019_assessment} ). For semiconductors, however, PGSL and PGSLr show little improvement and remain comparable to PGS and KGE2 despite the addition of two tuning parameters. We observed similar behavior when adding mGGA terms to KGE2 and LKT: while accuracy improved slightly, it required tuning coefficients and violated exact GE2 constraints. 
Since the primary objective of the present work is to develop a simple, transferable, and parameter-free semilocal functional that 
satisfies several known exact and physically motivated constraints,  
we do not pursue such extensions further.

PGSLr was specifically constructed to preserve the jellium response while incorporating parameter tuning. We argue that rather than introducing additional empirical terms, future improvements should focus on refining the semilocal description within the GGA framework.

\subsection{\label{sec:Result:level-4.1-RMARE} Assessment of Bulk Properties Using RMARE}

\begin{figure*}
    \centering
    \includegraphics[scale=0.5]{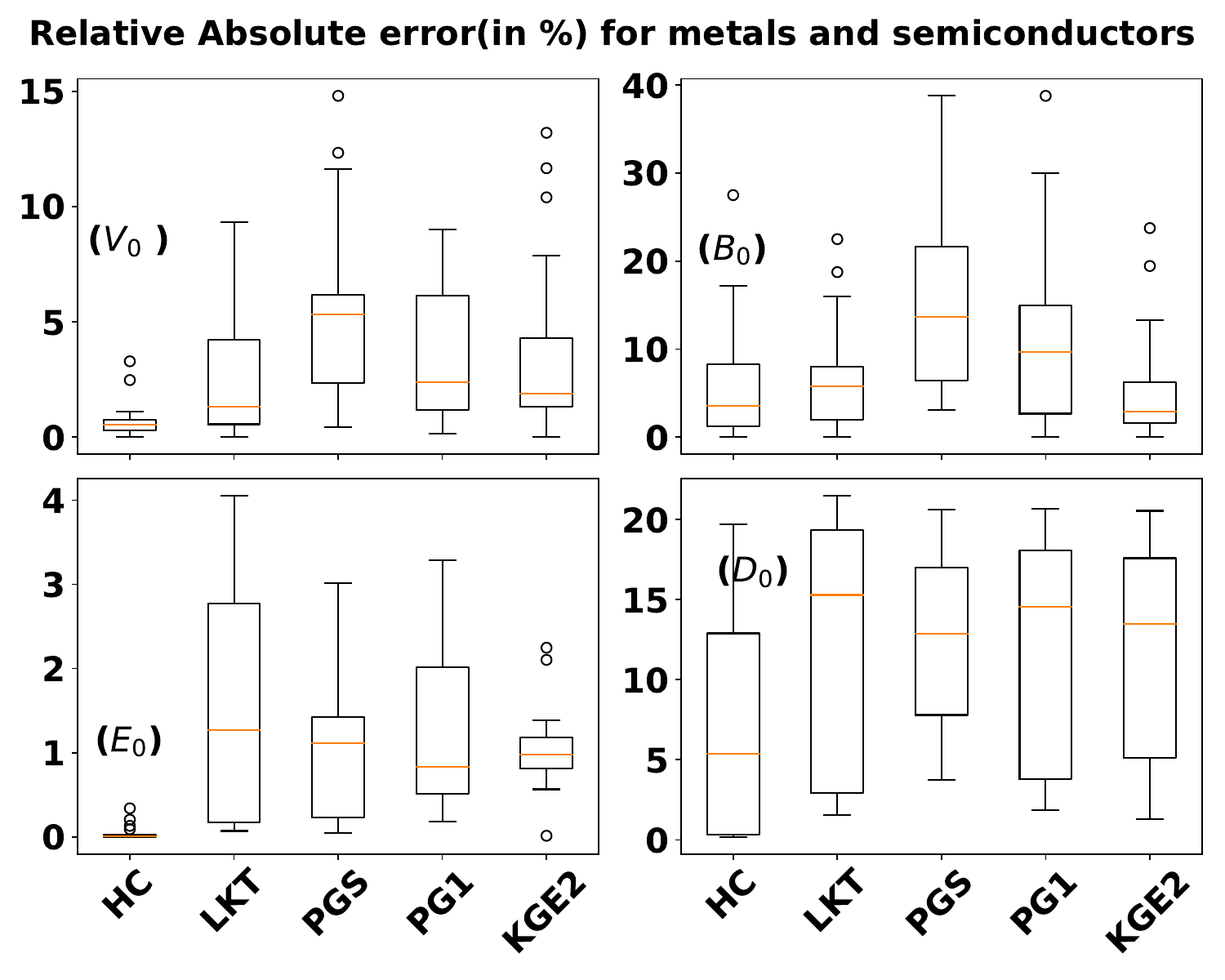}
    \caption{Box plots of the relative absolute error (in \%) of equilibrium volume ($V_{0}$), bulk modulus($B_{0}$), energy($E_0$), and density error  ($D_{0}$) for semiconductors and metals. The box plots summarize the overall distributions of the data reported in Tables I--IV of the Supporting Information \cite{support}. 
    }
    \label{fig:box_all}
\end{figure*}

\begin{table}
{\fontsize{8.5}{8.5}\selectfont
\caption{\label{tab:semi}
Comparison of KGE2 with state-of-the-art KEDFs. Equilibrium volume ($V_0$ in $\AA^3$) per atom, minimum energy ($E_0$ in eV/atom), and bulk modulus ($B_0$ in GPa) for some representative semiconducting and metallic systems. The boldface values represent the closest values to KS reference results.}
\begin{ruledtabular}
\begin{tabular}{ccccccc}
 elements & Functional &  & $V_0$    &    $E_0$  & $B_0$ & $D_{0}$ (\%) \\ \hline \hline
 CD-Si & KS \cite{MGP_2018} &  & 19.77 & -109.62 & 98 & \\
   & LKT &  & \textbf{19.77} & -107.09 & \textbf{96} & 14.41\\
   & \textcolor{black}{KGE2}  & $\alpha=1.48$ & 19.96 & -109.00 & 88.90 & 12.57 \\
   & \textcolor{black}{PGS}  & $\mu=1.48$ & 20.99 & -109.85 & 60 & \textbf{11.95}\\
   & \textcolor{black}{PG1} & $\mu=1.0$ & 20.27 & -107.863 &  82 & 14.52 \\
  \hline 
   AlAs & KS \cite{MGP_2018}&  & 21.80 & -117.89 & 80  &\\
   & LKT &  & 22.18 & -113.11 & 76 & 17.64\\
   & \textcolor{black}{KGE2}  & $\alpha=1.48$ & \textbf{21.99} & -115.24 &  \textbf{78} & 15.84\\
   & \textcolor{black}{PGS}  & $\mu=1.481$ & 23.15 & -116.25 & 66 &15.38\\
   & \textcolor{black}{PG1} & $\mu=1.0$ & 22.80   & -114.017  & 69 & \textbf{13.63} \\
  \hline 
  
   GaAs & KS \cite{MGP_2018}&  & 20.3 & -117.89 & 75 &\\
   & LKT &  & 18.64 & -118.08 & 95 &16.18\\
   & \textcolor{black}{KGE2} & $\alpha=1.48$ & 18.70 & -120.37 & 88 &14.36\\
   & \textcolor{black}{PGS} & $\mu=1.48$ & \textbf{19.89}  & -121.45 & \textbf{72} & \textbf{13.88}\\
   & \textcolor{black}{PG1} & $\mu=1.0$ & 19.01 & -119.044 & 92 & 14.6 \\
  \hline 
  
  InSb & KS \cite{MGP_2018}&  & 31.45 & -101.19 & 50 &\\
   & LKT &  & \textbf{31.25} & -98.55 & 49 & 20.28\\
   & \textcolor{black}{KGE2} & $\alpha=1.48$ & 30.90 & -100.19 & \textbf{49} &18.75\\
   & \textcolor{black}{PGS} & $\mu=1.48$ & 32.05 & -100.96 & 43 & \textbf{17.52} \\
   & \textcolor{black}{PG1} & $\mu=1.0$  & 31.90   & -99.240   & 44 & 18.84 \\
  \hline 
   Al-FCC & KS \cite{XMW_PhysRevB.100.205132} &  & 15.60 & -57.94 & 83 & \\
   & LKT &  & \textbf{16.82} & -58.04 & \textbf{88} &3.96 \\
   &\textcolor{black}{KGE2}& $\alpha=1.48$  & 17.06 & -58.60 & 72 & \textbf{1.31} \\
   &\textcolor{black}{PGS}& $\mu=1.48$  & 17.91 & -58.83 & 64 & 9.98 \\
   & \textcolor{black}{PG1} & $\mu=1.0$  & 16.97 & -58.242 & 81 & 4.77\\
  \hline 
     Li-BCC & KS \cite{XMW_PhysRevB.100.205132} &  & 18.82 & -7.59 & 16 & \\
    & LKT &  & \textbf{18.79} & -7.61 & \textbf{16} & \textbf{1.59} \\
   &\textcolor{black}{KGE2}& $\alpha=1.48$ & 18.10 & -7.66 & 17 &2.76 \\
   &\textcolor{black}{PGS}& $\mu=1.48$ & 17.80 & -7.69 & 17 & 3.75 \\
   & \textcolor{black}{PG1} & $\mu=1.0$ & 18.51 & -7.637 & 17 & 1.92 \\
  \hline 
\end{tabular}
\label{tab:Bulk-data}
\end{ruledtabular}
}
\end{table}

\begin{table*}
    \caption{ Relative median absolute error (RMARE) in (\%) of the various KEDFs for bulk solids. The boldface values represent the best values in semilocal level.}
    \centering
    \begin{tabular}{cccccccccc}
    \hline\hline 
    &   & SM & HC &  LKT & PGS & PG1 & KGE2 &  PGSL & PGSLr \\
    &  & ($A=0.2$) & ($\lambda=0.011$) & ($a=1.3$) & ($\mu=1.48$) & ($\mu=1$) & ($\alpha=1.48$) & ($\beta=0.25$) & ($\lambda=0.4$) \\
     & & & ($\beta=0.71$) & & & & & & ($\sigma=0.2$) \\   
  \hline\\
  RMARE (semi+metal)  &  & 6.825 & 1.174 & 6.984 & 7.771 & 7.40 & \textbf{5.822} & 6.984 & 6.370 \\
  RMARE (semi)  &  & 6.738 & 1.261 & 8.723 & \textbf{3.719} & 7.515 & 4.469 & 6.315 & 5.539 \\ 
  RMARE (metals) &  & 5.808 & 2.191 & \textbf{7.241} & 27.075 & 11.992 & 17.976 & 9.749 & 7.840  \\
    \end{tabular}
    \label{tab:RMARE-median}
\end{table*}

The accuracy of a kinetic energy density functional (KEDF) is typically evaluated based on its ability to reproduce both global and local properties. Global properties include structural and mechanical quantities such as equilibrium energy ($E_0$), bulk modulus ($B_0$), and equilibrium volume ($V_0$), while local properties are reflected in the accuracy of the density, potential, and kinetic energy density. In this work, we assess these aspects through four metrics: $V_0$, $E_0$, $B_0$, and the density error ($D_0$).

To systematically compare functional performance, we adopt the relative median absolute relative error (RMARE), as introduced in Ref.~\cite{PGSL_2018_doi:10.1021/acs.jpclett.8b01926}. The RMARE is defined as:
\begin{align}
    \text{RMARE} = \sum_{i} \frac{\text{MARE}_{i}}{ \frac{1}{2} ( \text{MARE}^{HC}_{i} + \text{MARE}^{SM}_{i}) }
    \label{eq:RMARE:solids}
\end{align}
where $i \in \{V_0, E_0, B_0, D_0\}$ and MARE denotes the median absolute relative error. Each component is normalized by the average error of two benchmark functionals: HC (optimized for semiconductors) and Smargiassi-Madden (SM)~\cite{SM_1993} (best for metals). To ensure a balanced assessment, the dataset consists of 10 semiconductors and 9 metals. The reference values of $V_0$, $E_0$, and $B_0$ for HC and SM are taken from Refs.~\cite{HC,KGAP_2018}. For the calculation of $D_0$, HC is evaluated using the recommended parameters $\lambda = 0.01177$, $\alpha=1.952$, and $\beta=0.7143$,
while SM is used with its standard internal parameter $A = 0.2$.

Figure~\ref{fig:box_all} and Table~\ref{tab:RMARE-median} summarize the RMARE statistics for each functional over the combined metals and semiconductors dataset. The box plots illustrate the distribution of errors, with quartiles ($Q1$ and $Q3$) defining the box boundaries, whiskers extending to the non-outlier extrema, and outliers marked individually. The yellow band indicates the median error, while the box height 
represents the interquartile range (IQR). 
The complete dataset used to generate  Fig.~\ref{fig:box_all} and Table~\ref{tab:RMARE-median} is provided in the Supporting Information~\cite{support}, with results for metals and semiconductors reported in Tables III and IV, respectively. 
Representative values for selected systems are also presented in Table~\ref{tab:Bulk-data}.

Among the semilocal GGA-level functionals, KGE2 exhibits the best overall performance for the combined set of metals and semiconductors, achieving the lowest RMARE (5.82\%), outperforming LKT (6.98\%), PG1 (7.40\%), and PGS (7.71\%). Even the mGGA extensions of PG-class functionals fail to surpass KGE2 in accuracy despite their higher computational cost. This result supports our goal of constructing a simple, parameter-free, generally accurate, and computationally efficient KEDF.

For semiconductors, PGS yields the most accurate equilibrium energy ($E_0$) and the lowest $D_0$ among the semilocal functionals, with an RMARE of 3.71\%. However, its performance for $V_0$ and $B_0$ are much less satisfactory, which indicates deviation from correct response function in high-$s$ regime. Here we also note that although in low-s limit PGS and KGE2 become GE2, the accuracy of high-$s$ limit (inhomogeneous density) is purely model dependent where KGE2 clearly is better than PGS to get correct $V_0$ and $B_0$. By construction, KGE2 mimics LKT in high-$s$ regions from Fig.\ref{fig:F_theta_enhance} and deviates from PGS. Thus KGE2 in interstitial region mimics PGS to yield a good density while capturing the accurate bulk properties, giving a balanced performance in overall aspects.   PG1, constructed by tuning the $\mu$ parameter, improves $V_0$ and $B_0$ but worsens $E_0$ and $D_0$, raising the overall RMARE to 7.51\%. This highlights the drawbacks of parameter tuning, which KGE2 avoids by construction. With an RMARE of 4.47\%, KGE2 outperforms even PGS and PGSLr, the latter of which includes mGGA terms. The relatively poor performance of PGSLr in semiconductors suggests that its parameters ($\beta = 0.25$, $\lambda = 0.4$, $\sigma = 0.2$) deviate from optimal behavior for localized systems. LKT performs the worst in this category, despite relatively stable values for $V_0$ and $B_0$ across both material types.

For metals, however, the trend is reversed.
Here, LKT performs best among all semilocal and mGGA functionals, with an RMARE of 7.24\%, followed by PGSLr and PGSL. KGE2 performs comparatively poorly, while PGS shows the highest error, nearly four times that of LKT. The addition of mGGA terms in PGSLr improves both structural and density predictions for metals, suggesting that it better captures metallic response behavior. The sharp decline of the enhancement factor $F_\theta(s)$ in PG at large $s$ effectively suppresses the TF component, leading to an overemphasis on the vW term, a behavior akin to a bosonic system. This imbalance can be corrected either by reducing the steepness of the slope ($\mu: 1.48 \to 1.0$) or by incorporating mGGA terms to compensate. This rationale also explains why the GE4 coefficient $\beta = 8/81$ is increased to $\beta = 1/4$ in PGSL-type functionals. PGSLr also benefits from a well-behaved response function, as illustrated in Fig.~3 of Ref.~\cite{PGSL_2018_doi:10.1021/acs.jpclett.8b01926}. However, the increased parameter count and higher computational cost, along with poor convergence behavior, especially for finite systems, limit their practicality as general-purpose OF-KEDFs.

From this analysis, we conclude that metallic systems benefit from a slowly decaying $F_\theta(s)$ in the low-$s$ regime, whereas semiconductors perform better when $F_\theta(s)$ decays rapidly at high $s$. The transition point for this change in behavior is system-specific. In practice, tuning parameters in enhancement factors indirectly controls this inflection point. Ideally, a semilocal KEDF would treat this crossover as a spatially varying, nonlocal feature, modulating the TF contribution based on the local environment.


\subsection{\label{sec:Result:level-4-Elastic}Elastic Properties}
A precise description of the density response function is essential for the construction of a high-quality KEDF 
\cite{10.1063/5.0146167}. Mechanical properties 
provide a stringent test of a KEDF’s performance, as they are directly linked to the system’s linear response behavior~\cite{PhysRevB.85.045126,Elastic_carter}. When a crystal lattice is deformed, the external potential is perturbed, which induces changes in the electron density. A reliable KEDF must correctly reproduce the resulting stress response through its dependence on the density.

The relationship between stress ($\boldsymbol{\sigma}$) and strain ($\boldsymbol{\epsilon}$) is given by $\sigma_i = \sum_j C_{ij} \epsilon_j$, where $C_{ij}$ is the elastic stiffness tensor~\cite{JonesAndGunnarsson}.

\begin{table}[h!]
\caption{The elastic constants: C11, C12 and  triaxial shear modulus (C44) for some semiconductors calculated by KSDFT and OFDFT with different KEDFs. 
For HC we have used optimal choices of the parameters mentioned in \cite{HC}.
(* denotes unstable structure.) 
}
\fontsize{8}{10}\selectfont
\begin{tabular}{ ccccccc }  
  \hline 
  elements & Functional &  &  & $C_{11}(GPa)$ & $C_{12}(GPa)$ & $C_{44}(GPa)$  \\
  \hline\hline
   Si-CD  & KS (\cite{Elastic_carter}) &  &  & 162 & 65  & 76  \\
    & HC &  &  & 101 & 92 & 81 \\
    & KGE2 *&  &  & 58 & 122 & 35 \\
    & PG1 *&  &  & 2 & 301 & -142  \\
  \hline 
    GaP & KS &  &  & 141 & 67 & 67  \\
    & HC *&  &  & 12 & 89 & -39  \\
    & KGE2 &  &  & 96 & 31 & 19 \\
    & PG1 &  &  & 349 & -4 & 54 \\
    \hline
    GaSb & KS &  &  & 86 & 40 & 41 \\
    & HC &  &  & 51 & 49 & 43 \\
    & KGE2 *&  &  & 20 & 63 & 26 \\
    & PG1 *&  &  & 248 & 58 & -44 \\
    \hline
    InAs & KS &  &  & 84 & 49 & 36 \\
    & HC *&  &  & 34 & 44 & 22 \\
    & KGE2 *&  &  & 29 & 61 & 79 \\
    & PG1 *&  &  & -176 & 131 & 120 \\
    \hline
    InP & KS &  &  & 99 & 58 & 42 \\
    & HC *&  &  & 31 & 94 & 37 \\
    & KGE2 &  &  & 132 & 60 & 92 \\
    & PG1 &  &  & 1719 & 86 & 1051 \\
    \hline
\end{tabular}
    \label{tab:Elastic}
\end{table}
For cubic crystals, symmetry reduces the number of independent components to three: $C_{11}$, $C_{12}$, and $C_{44}$. The elastic constants were obtained by applying small strains to the equilibrium crystal structures. We compute $C_{44}$ by applying a triaxial shear strain $\boldsymbol{\epsilon} = (0,0,0,\delta,\delta,\delta)$, and $C_{11}$ and $C_{12}$ using a biaxial strain $\boldsymbol{\epsilon} = (\delta,\delta,0,0,0,0)$. 
 The bulk modulus is related to the elastic constants through $B_0 = (C_{11} + 2C_{12})/3$ and $C_{44}$ is obtained from the strain-energy relation $\Delta E / V_0 = \frac{3}{2} C_{44} \delta^2$.
 
For the cubic systems considered here, a reliable KEDF should predict elastic constants that satisfy the Born--Huang mechanical stability criteria: 
\begin{enumerate}
    \item $C_{11} - C_{12} > 0$,
    \item $C_{11} + 2C_{12} > 0$,
    \item $C_{44} > 0$.
\end{enumerate}

Selected results are presented in Table~\ref{tab:Elastic}, while the complete dataset is provided in the Supporting Information (see Table V).
The KSDFT reference values satisfy the mechanical stability criteria for all systems studied. 
In contrast, several OFDFT functionals, including advanced nonlocal KEDFs such as HC and KGAP, violate one or more stability conditions, reflecting the sensitivity of elastic properties to the underlying density-response description. 
Not all OFDFT predictions satisfy these stability criteria, indicating substantial room for improvement in KEDFs for elastic properties.
Our proposed semilocal functional, KGE2, performs competitively with complex NL-KEDFs such as HC and JGM \cite{JGM-10.1063/5.0204957}. Despite structural similarities between PG1 and KGE2, PG1 yields substantial errors in elastic constants, while TFvW performs the worst across all metrics.

Huang and Carter previously reported~\cite{Elastic_carter} that WGC-class functional fails to accurately predict elastic constants for the Si-CD phase. For GaP and InP, both PG1 and KGE2 satisfy the Born–Huang stability criteria, whereas several of the other tested KEDFs violate one or more criteria.
In contrast, MGP fails to accurately reproduce the elastic tensor, often yielding negative $C_{11}$ and $C_{12}$ values~\cite{JGM-10.1063/5.0204957}. Other functionals, including WT, SM, and TFvW, also perform poorly in this regard (see Table V of the Supporting Information).

%

\subsection{\label{sec:Result:level-5-clusters} Application of KGE2 on finite systems: Clusters}

\begin{figure}
    \centering
    \includegraphics[scale=0.4]{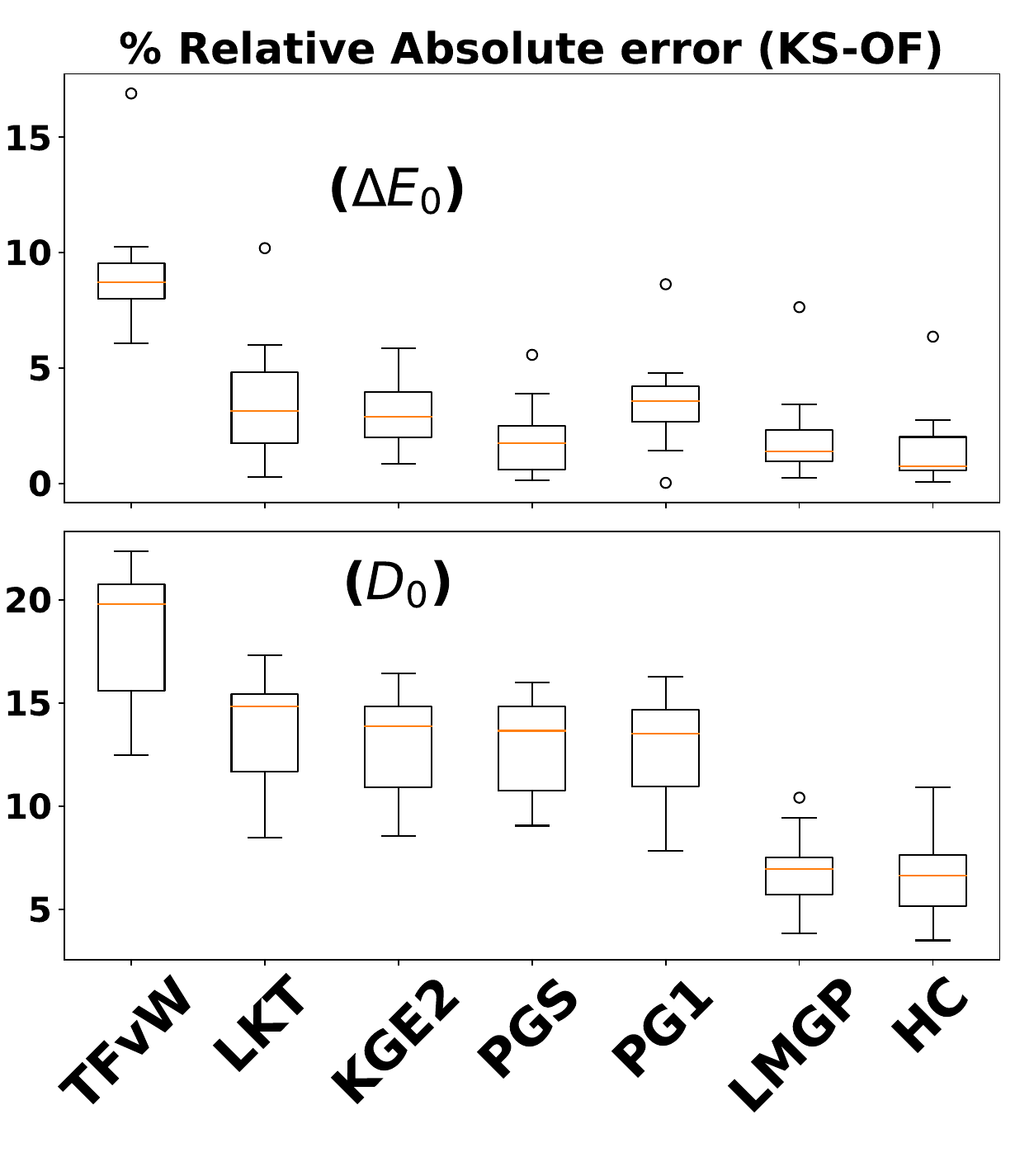}
    \caption{Boxplots of the absolute error of energy($E_0$) in eV and density-error ($D_{0}$) for the cluster systems considered in this work. The boxplots summarize the overall distribution of the data reported in Tables VI and Table VII of the Supporting Information\cite{support}. 
    }
    \label{fig:Boxplot_cluser}
\end{figure}

\begin{figure}
    \centering
    \includegraphics[scale=0.3]{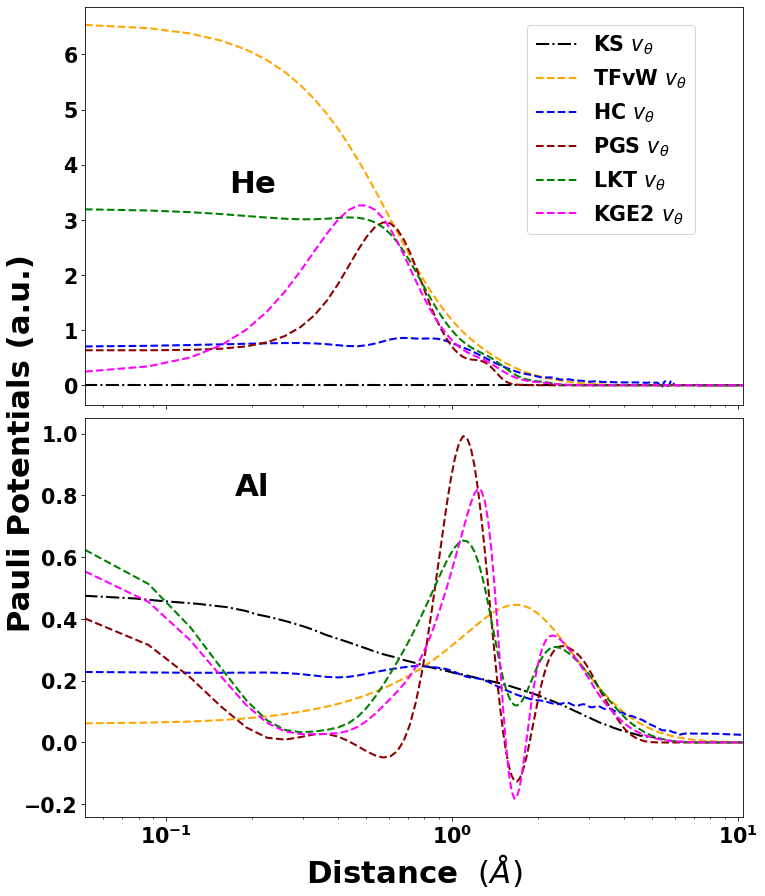}
    \caption{Numerical Pauli potential for the He atom (upper panel) and Al atom (lower panel) calculated using the various KEDFs considered in this work.}
    \label{fig:Pauli-pot-He}
\end{figure}

\begin{figure}
    \centering
    \includegraphics[scale=0.3]{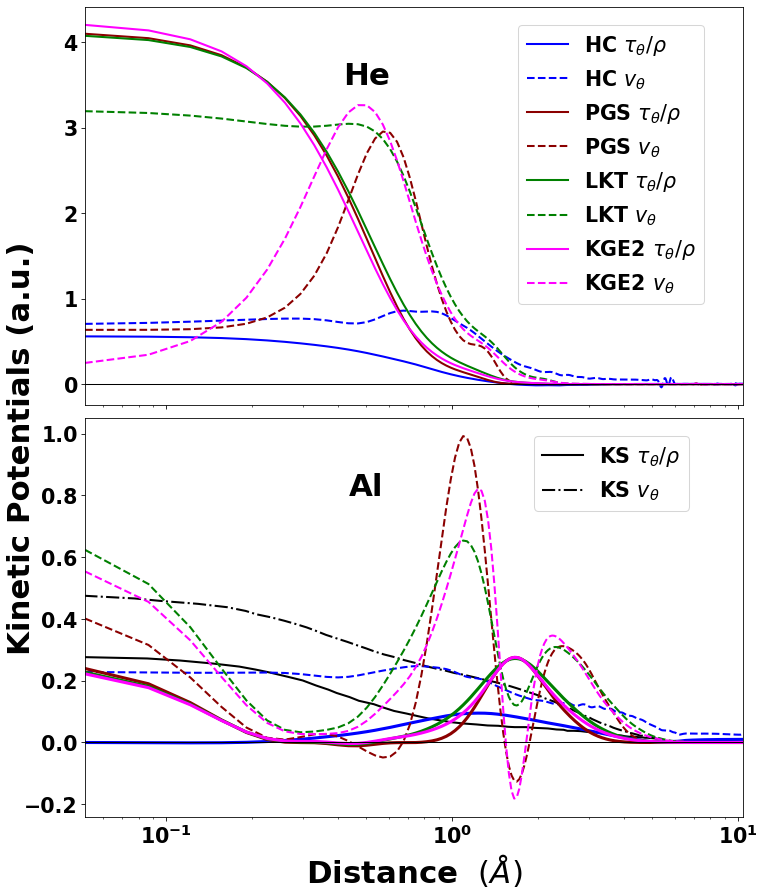}
    \caption{Numerical kinetic energy densities for the He atom (upper panel) and Al atom (lower panel) calculated using the various KEDFs considered in this work. }
    \label{fig:Pauli-pot-Al}
\end{figure}

To assess KEDF performance on finite systems, we examine four key metrics: (1) electron density, (2) Pauli potential, (3) kinetic energy density, and (4) total energy. Results for these quantities are presented in Figs. \ref{fig:Boxplot_cluser}, \ref{fig:Pauli-pot-He}, and \ref{fig:Pauli-pot-Al}. Although absolute total energies suffer from significant error cancellation among different Hamiltonian components and thus carry limited physical weight, they are included for 
completeness and direct comparison with previous studies. We limit our study of semilocal KEDFs to the GGA level because mGGA terms often amplify numerical noise in low‐density regions, leading to convergence difficulties in PAW based codes such as DFTpy. In existing semilocal functionals, the mGGA contribution serves primarily as a corrective term to the GGA baseline, which accounts for the dominant fraction  of the kinetic energy. Accordingly, our focus here is on constructing an accurate GGA component, with the expectation that  subsequent inclusion of mGGA corrections may further improve the density description. Finite systems, particularly isolated atoms, are especially valuable test cases because they eliminate spurious density overlap and thereby 
exposing the intrinsic behavior of a functional.

For the global error indicator, we have modified the definition of RMARE from Eq.\ref{eq:RMARE:solids} as 
\begin{align}
    \text{RMARE} =  \sum_{i} \frac{\text{MARE}_{i}}{\frac{1}{2}[ \text{MARE}^{HC}_{i} + \text{MARE}^{LMGP}_{i}] }
    \label{eq:RMARE:clusters}
\end{align}
where $i \in \{\Delta E, D_{0} \}$ with $\Delta E$ are same as defined in Sec.\ref{sec:Result:level-4.1-RMARE}. 
The benchmark set consists of the 17 cluster systems described in 
Sec.\ref{sec:Result:level-2-comp.details}.


\begin{table*}[htbp]
    \centering
    \caption{Relative median absolute relative errors (RMAREs) of the various KEDFs  for the cluster test set.}
    \label{tab:RMARE}
    \renewcommand{\arraystretch}{1.3} 
    \setlength{\tabcolsep}{10pt} 
    
    \begin{adjustbox}{max width=\textwidth}
    \begin{tabular}{lcccccc}
        \toprule
        \textbf{Functional} & TFvW & LMGP ($A=0.2$) & HC ($\lambda=0.01177$) & LKT & KGE2 & PGS \\
        \midrule
        RMARE  & 11.10 & 2.33 & 1.66 & 5.12 & 4.76 & 3.64 \\
        \bottomrule
    \end{tabular}
    \end{adjustbox}
\end{table*}

Figure~\ref{fig:Boxplot_cluser} presents boxplots 
of the relative absolute error of energy and density, analogous to 
Fig.~\ref{fig:box_all}. 
The interpretation of the boxplots here remains the same as explained in Sec. \ref{sec:Result:level-4.1-RMARE}. We benchmark KGE2 against TFvW, PGS, PG1, and LKT, as well as the best NL-KEDFs: 
HC and LMGP, employing the parameters specified in Sec.~\ref{Sec:Result:level-1}. 
In Table~\ref{tab:RMARE} we report the RMARE values for various semilocal and nonlocal KEDFs.
The total energies, and density errors ($D_0$) used in the RMARE analysis are provided in the Supporting Information (see Tables VI and VII). 
The results further ascertains our claim that performance of KGE2 lies between PGS and LKT. However, we note that among the semilocals KEDFs considered in this work,  TFvW performs poorly in describing energies and densities.
It is noteworthy that PG1 also performed  poorly for clusters, likely because its parameter $\mu=1$ was optimized primarily for bulk solids. In contrast, PGS 
performs significantly better for finite systems, consistent with previous observations for atoms and jellium clusters \cite{GSE2_2019_Lucian_PhysRevB.99.155137}. 


Regarding non-local Pauli functionals, 
previous studies have shown that the Pauli positivity condition can be violated by many nonlocal KEDF approximations, particularly in low-density or isolated-atom limits  \cite{Shao(2024)-effective-WT-PhysRevB.110.085129}. 
In WT-type nonlocal functionals, this behavior is closely related to the Blanc–Cancès variational instability and to the violation of the exact uniform density-scaling relation. In particular, the WT functional does not satisfy the exact scaling condition ($T_s[n_\lambda]=\lambda^2T_s[n]$).
These difficulties are less severe for the semilocal class of functionals considered here, since the Pauli-potential constraints discussed in Sec.~\ref{sec:Theory:level-3-pauli} are theoretically satisfied by construction. However, PG-type functionals may still exhibit apparent violations in some cases because of numerical noise. 

As shown in Fig.~\ref{fig:Pauli-pot-He}, the helium and aluminum atoms serve as representative test systems for assessing the performance of the investigated KEDFs through their corresponding Pauli potentials. The results show that almost all the tested KEDFs satisfy the Pauli positivity condition, with PGS in Fig.~\ref{fig:Pauli-pot-Al} being an exception attributed to strong numerical noise.

The violation of constraint-(d): $v_{\theta}({\bf{r}}) \ge t_{\theta}(\textbf{r}) $, in the near nuclear region is a major indicator that showcases the drawback of simplistic description of semilocals which on the other hand nonlocals satisfy with ease.
It is also interesting to notice how, in the long range, the fall of the Pauli potential is slower than all the semilocals. And just after the plateau region, all the semilocals converge with HC. This brings us to the question how should be the exact decay rate of Pauli potential, and in the work of \cite{Exact-pauli-Levy_PhysRevA.38.625, GSE2_2019_Lucian_PhysRevB.99.155137}, for spherical-symmetry atoms the Pauli potential in the tailing end decays as $\frac{1}{r^2}$, in fact, the exact expression given as: $v_{\theta} \to \frac{l(l+1)}{2} \frac{1}{r^2};$ where l is the azimuthal quantum number. This decay rate is obeyed by almost all of the functionals correctly when away from the nucleus.

We report the median of absolute error (MAE) in energy of about 1.15 eV for PGS, which is about 1.74\% error (MARE) is the least error followed by KGE2 and LKT, respectively, with percentage errors 3.15\% and 3.42\% in Fig.~\ref{fig:Boxplot_cluser}. And with a relative error in density of 12.89\% for PGS, 13.19\% for KGE2 and 13.65\% for LKT. Thus the PGS functional marginally outperforms KGE2 and LKT for the finite system  considered here. This observation is in par with our analysis before in section.~\ref{Sec:Result:level-1}, where PGS outperformed in density and energy for localized systems. The steep slope of $F_{\theta}(s)$ in Fig.\ref{fig:F_theta_enhance} decays the TF quickly, highlighting the vW term which is more suited to describe the inhomogeneous cluster densities. Of course, the density-dependent nonlocal kernels like HC and LMGP outperform all the semilocals by a large margin. Note that this error in energy might seem huge as compared to bulk systems where the \% error of energy of HC is typically 0.46\% and 8.3\% for energy and density respectively (see table.1 of \cite{PGSL_2018_doi:10.1021/acs.jpclett.8b01926}). This systematic large MARE for the local system is due to the presence of a large vacuum present in the simulation box \cite{PROFESS:software}, and we have also found this vacuum responsible for the slow convergence of many semilocal functionals (especially PG-class). 

Though the pure non-empirical nature of semilocal as compared to nonlocals gives an extra edge for the predictability of unknown materials, however, the Pauli potential of all the semilocals behaves similarly and poorly. We believe that the uncontrollable oscillation of the Pauli potential for all the semilocal KEDFs  hints towards some major drawback, resolving which can result in better accuracy of one-point kinetic energy functionals in general.

\section{\label{sec5-conclusion}Conclusion}


In this work, we have developed KGE2, a parameter-free semilocal kinetic energy density functional constructed within the generalized gradient approximation (GGA) framework. The functional is  constructed using a constraint-based approach, with particular emphasis on recovery of the exact second-order gradient expansion (GE2) in the slowly varying density limit while retaining the simplicity 
of a semilocal form.

KGE2 satisfies several exact and physically motivated constraints, including 
recovery of the GE2 limit, the corresponding
long-wavelength response behavior, and the uniform coordinate
scaling relations considered here. 

Benchmark calculations were carried out for metals, semiconductors, and finite clusters within the BLPS pseudopotential framework. Our benchmarks demonstrate that KGE2 performs comparably to state-of-the-art semilocal KEDFs in predicting structural and mechanical properties of both metals and semiconductors. Notably, KGE2 consistently yields balanced accuracy across materials classes, outperforming PG-class for metals and LKT for semiconductors. 
Although meta-GGA-level functionals like PGSL and PGSLr show significantly good results for specific material types, they involve parameter fitting and incur a higher computational cost. In contrast, KGE2 maintains parameter-free construction and still surpasses PGSL and PGSLr in average performance for solids. For finite systems, especially small clusters, nonlocal functionals still outperform semilocal ones by a significant margin, though the performance gap narrows with increasing cluster size. 
Though semilocal KEDFs satisfy all the Pauli constraints by construction, we have shown that some important constraints do fail numerically. Investigating these could bring new insights to drawbacks of semilocal class of functionals.

Overall, KGE2 provides a physically motivated, simple, computationally inexpensive, and parameter-free KEDF 
with 
balanced performance across the both bulk and finite systems considered here. 
These results support further assessment of KGE2 as a transferable semilocal KEDF for orbital-free DFT applications. 

\section*{Data Availability \label{sec6:data_availability_statement}}
The data supporting the findings of this article are available freely in supplementary materials~\cite{support}. 
The modified version of DFTpy containing the KGE2 implementation used in this work is publicly available on Zenodo~\cite{DFTpyKGE2}. 

\section*{Acknowledgments}
%
The authors acknowledge partial financial and computational support from the National Institute of Science Education and Research (NISER). The author thanks physics department of IIT Kanpur for partial funding.


\appendix

\section{\label{App:Pauli-pot} Pauli Potential of semilocals:}

Any generic semilocal functional can be written in the following form:
\begin{align}
    T_{\theta} = \int d\textbf{r} \tau_{TF}(\rho)F_{\theta}(s) \nonumber
\end{align}
And the Pauli potential as defined in Sec.\ref{sec1:Introduction:level1},
\begin{align}
    v_{\theta}(\textbf{r}) &= \frac{\delta}{\delta\rho(\textbf{r}')} \int d\textbf{r} \tau_{TF}(\rho)F_{\theta}(s) \nonumber\\
    &= \frac{\partial}{\partial\rho} (\tau_{TF}F_{\theta}) - \nabla \frac{\partial}{\partial(\nabla(\rho)} (\tau_{TF}F_{\theta}) \nonumber \\
    &= \frac{5}{3}C_{TF} \rho^{2/3}F_{\theta}(s) + C_{TF} \rho^{5/3}\frac{\partial F_{\theta}}{\partial\rho} - \nabla [\tau_{TF}\frac{\partial F_{\theta}}{\partial(\nabla\rho)}] \nonumber \\
    &= \frac{5}{3} C_{TF} \rho^{2/3} F_{\theta} + C_{TF} \rho^{5/3} \frac{\partial F_{\theta}}{\partial s}.\frac{\partial s}{\partial \rho} - \nabla \Big[C_{TF} \rho^{5/3} \frac{\partial F_{\theta}}{\partial s} \frac{\partial s}{\partial \rho} \Big] \nonumber \\
    v_{\theta}(r) &= \frac{5}{3}C_{TF} \rho^{2/3}F_{\theta}(s) - \frac{4}{3}C_{TF}\rho^{2/3} s (\frac{\partial F_{\theta}}{\partial s}) - \frac{C_{TF}}{c_{s}}\nabla[\rho^{1/3}\frac{\partial F_{\theta}}{\partial s}]  
\end{align}

where we have used, $\frac{\partial F_{\theta}}{\partial\rho} = \frac{\partial F_{\theta}}{\partial s}.\frac{\partial s}{\partial \rho}=-\frac{4}{3}\frac{s}{\rho}.\frac{\partial F_{\theta}}{\partial s}$ and $\frac{\partial F_{\theta}}{\partial(\nabla\rho)} =\frac{1}{c_{s}\rho^{4/3}}.\frac{\partial F_{\theta}}{\partial s}$, 

We can see that putting $F_{\theta}=1$ readily gives us TF potential. Now plugging the form of Eq.\ref{eq:F_tr02} we get 

\begin{align}
    v^{KGE2}_{\theta}(\textbf{r}) = &\frac{1}{3} \frac{C_{TF} \rho^{2/3}(\textbf{r})}{(1+\alpha s^{2})^{2}} [ 5+11\alpha s^{2} ]  + \frac{2}{3} \frac{\alpha C_{TF}\rho^{2/3}(\textbf{r})}{(1+\alpha s^{2})^{3}}  \nonumber \\
     & ~ \times\Big[ -3s^{2} + 13\alpha s^{4} + \frac{3\nabla^{2}\rho(\textbf{r})}{c_s^{2}\rho^{5/3}(\textbf{r})} \{1-9\alpha s^{2} \}  \Big] 
     \label{eq:v_theta_KGE2}
\end{align}

\section{\label{App:Exact-conditions} Exact conditions of $v_{\theta}$}


\subsection{\label{App:Exact-Scaling} Scaling conditions}
Here we show how scaling condition is satisfied by KGE2 functional. The scaling conditions are given by: $$v_{\theta}[\rho_{\lambda}](\textbf{r}=\textbf{r}_{0}) = \lambda^{2} v_{\theta}[\rho](\textbf{r}=\lambda \textbf{r}) $$ where $\rho_{\lambda}=\lambda^{3}\rho(\lambda \textbf{r})$.

\begin{align}
    v_{\theta}(\textbf{r}) = \frac{1}{3} & \frac{C_{TF} \rho^{2/3}(\textbf{r})}{(1+\alpha s^{2})^{2}} [ 5+11\alpha s^{2} ]  + \frac{2}{3} \frac{\alpha C_{TF}\rho^{2/3}(\textbf{r})}{(1+\alpha s^{2})^{3}}  \nonumber \\ 
     & \times\Big[ -3s^{2} + 13\alpha s^{4} + \frac{3\nabla^{2}\rho(\textbf{r})}{c_s^{2}\rho^{5/3}(\textbf{r})} \{1-9\alpha s^{2} \}  \Big] \nonumber
\end{align}
Let's check how $s(\rho,\nabla\rho)$ behaves under this scaling.
\begin{align}
    s(\rho_{\lambda}) &= \frac{\nabla(\lambda^{3}\rho(\lambda\textbf{r}) )}{c_s \lambda^{4} \rho^{4/3}(\lambda\textbf{r}) } \nonumber \\
    &= \frac{\nabla_{\lambda}\rho(\lambda\textbf{r})}{c_{s}\rho_{\lambda}^{4/3}(\textbf{r})} = s(\lambda \textbf{r}) \nonumber \\
    \text{Thus, }\quad s(\rho_{\lambda})\to & s(\lambda\textbf{r}); \quad s^{2}(\rho_{\lambda})\to s^{2}(\lambda\textbf{r})
\end{align}
Similarly it can be shown that, $\frac{\nabla^{2}\rho(\textbf{r})}{\rho^{5/3}(\textbf{r})}\to \frac{\nabla^{2}\rho(\lambda\textbf{r})}{\rho^{5/3}(\lambda\textbf{r})}$. Using these it can be shown that,

\begin{align}
    v_{\theta}^{KGE2}[\rho_{\lambda}](r=r_{0}) &= \lambda^{2} v_{\theta}^{KGE2}[\rho](r=\lambda r_{0})
    \label{eq:scaling_v_theta}
\end{align}

\section{\label{App:Linear_Response} Response function for GGA functionals}

\subsection{ Method 1: Perturbing the kinetic potential }
The following steps are taken in order :
\begin{itemize}
    \item Calculate (kinetic)potential from KE expression.
    \item Perturb the density around $r=0$, and plug in potential.
    \item Take functional derivative.
\end{itemize}
Consider a small perturbation in the uniform electron gas (UEG) density, $k_{F}=(3\pi^2 \rho_{0})^{1/3}$, $\rho=\rho_0 + \rho_q e^{iq.r}$ at $r=0$ we get $\rho=\rho_0+\rho_q$, $\nabla\rho=iq\rho_q; ~ \nabla^2\rho=-q^2\rho_q$, which implies:
\begin{align}
    s=\frac{1}{2(3\pi^2)^{1/3}} \frac{iq\rho_q}{(\rho_0+\rho_q)^{4/3}}
\end{align}


\subsubsection{Response of \textbf{vW Functional}: }

\begin{align}
    T^{vW} &= \int \frac{1}{8} \frac{|\nabla\rho|^2}{\rho}~dr;  \nonumber \\ v^{vW}&= -\frac{1}{8} \frac{(\nabla\rho)^2}{\rho^2} + \frac{1}{4}\frac{\nabla^2\rho}{\rho} \quad = \frac{q^2}{8}(2\frac{\rho_q^2}{\rho^2}  - 2\frac{1}{\rho})
\end{align}
Now we take the limit $\rho_0 >> \rho_q$ and $\delta\rho=\rho_q\to0$ for small perturbation.
\begin{align}
    \frac{\delta v^{vW}}{\delta \rho}  & = - \frac{q^2}{4\rho_0};  \nonumber\\
    \frac{1}{\chi^{vW}} &= \frac{\pi^2}{k_{F}} 3\eta^2 
\end{align}
In the last step we have used the following: $\rho_0=\frac{k_{F}^3}{3\pi^2}$, $\eta=\frac{q}{2k_F}$

\subsubsection{Response of \textbf{KGE2 functional}: }

\begin{align}
    v_{\theta}^{KGE2}(\textbf{r}) = \frac{1}{3} & \frac{C_{TF} \rho^{2/3}(\textbf{r})}{(1+\alpha s^{2})^{2}} [ 5+11\alpha s^{2} ]  + \frac{2}{3} \frac{\alpha C_{TF}\rho^{2/3}(\textbf{r})}{(1+\alpha s^{2})^{3}}  \nonumber \\ 
     & \times\Big[ -3s^{2} + 13\alpha s^{4} + \frac{3\nabla^{2}\rho(\textbf{r})}{c_s^{2}\rho^{5/3}(\textbf{r})} \{1-9\alpha s^{2} \}  \Big] \nonumber
\end{align}
Where $C_{TF}=\frac{3}{10}(3\pi^2)^{2/3}$; $c_s=2(3\pi^2)^{1/3}$.
\\
Lets work with \textbf{first term }($v_\theta^{KGE2,1}$):
\\
\begin{align}
    v_\theta^{KGE2,1} &= \frac{1}{3}C_{TF} \frac{\rho^{2/3}}{(1+\alpha s^2)^2} [ 1+\frac{11}{5}\alpha s^2 ] \nonumber \\
    \frac{1}{(1+\alpha s^2)^2} &\approx 1 - 2 \alpha s + ..  
\end{align}

\begin{align}
    v_{\theta}^{KGE2,1} &= \frac{5}{3}C_{TF}\rho^{2/3} \frac{1}{(1+\alpha s^2)} [1 - \frac{11}{5}\frac{\alpha}{4(3\pi^2)^{2/3}} \frac{-q^2\rho_q^2}{\rho^{8/3}} ] \nonumber \\
    &= \frac{5}{3}C_{TF}\rho^{2/3} (1 + 2 \alpha' \frac{q^2 \rho_k^2}{\rho^{8/3}}) ( 1  - \frac{11}{5}\alpha' \frac{q^2\rho_q^2}{\rho^{8/3}} + \mathcal{O}(\rho_q^4) ) \nonumber\\
    &= \frac{5}{3}C_{TF}\rho_0^{2/3}(1 - \frac{1}{5} \alpha' \frac{q^2~\rho_q^2}{\rho_0^{8/3}} + \mathcal{O}(\rho_q^3))
\end{align}

\begin{align}
    (\chi^{-1})^{1st} &= \frac{\delta }{\delta \rho}v_{1}^{KGE2} = \frac{5}{3} C_{TF}\rho_0^{2/3}(\frac{2}{3\rho_0^{1/3}}-\frac{6}{5}\frac{\alpha'q^2}{\rho_0^{8/3}}\rho_q ) \nonumber\\
    &= \frac{\pi^{2}}{k_{F}} - \frac{3}{5}\alpha\eta^2\frac{\rho_q}{\rho_0} \qquad \text{where, } \eta=\frac{q}{2k_{F}} \nonumber \\
    (\chi^{-1})^{1st} &= \chi^{-1}_{TF}-\frac{3}{5}\alpha\eta^{2}\frac{\rho_q}{\rho_0}
\end{align}
\\
Now with \textbf{second and third term}:
\\
\begin{align}
    v_{\theta}^{KGE2,2,3} &= \frac{2}{3} C_{TF}\alpha\rho^{2/3}\frac{1}{(1+\alpha s^2)^3}\{ -3s^2 + 13 s^4\} \nonumber \\
    &= \frac{2}{3} C_{TF}\alpha\rho^{2/3} (1-3\alpha s^2 + \mathcal{O}(s^4) ) (-3s^2 + \mathcal{O}(s^4) ) \nonumber \\
    &= \frac{2}{3} C_{TF}\alpha\rho^{2/3}_0 (-3s^2 + \mathcal{O}(s^4)) \nonumber
\end{align}

Now, $s^2\rho_q^2=-\frac{q^2}{4(3\pi^2)^{2/3}} \frac{\rho_q^4}{\rho_0^{8/3}}$; $\frac{\partial(s^2\rho_q^2)}{\partial\rho_q}=- \frac{6 q^2}{4(3\pi^2)^{2/3}}\frac{\rho_q^3}{\rho_0^{8/3}}$; $\frac{\partial (s^4\rho_q^2)}{\partial\rho_q}=\frac{6 q^4}{(4(3\pi^2))^{2}}\frac{\rho_q^{5}}{\rho^{16/3}}$. And we discard all the terms of order above $\mathcal{O}(\rho_q^2)$. With these simplifications we have final form putting expression of $s^2$ :

\begin{align}
    v_{\theta}^{KGE2,2,3} &= \frac{3}{20} \alpha q^2 (\frac{\rho_q}{\rho_0})^2 (1-\frac{8}{3}\frac{\rho_k}{\rho_0})
\end{align}

\begin{align}
    \frac{\partial v^{KGE2,2,3}_{\theta}}{\partial\rho_q} &= \frac{3}{20} \alpha q^2 (\frac{\rho_q}{\rho_0})^2 (1-\frac{8}{3}3\frac{\rho_q^2}{\rho_0}) \\
    (\chi^{-1})^{2,3} &= \frac{3}{10} \frac{\alpha q^2}{\rho_0^2} [1 - 8\frac{\rho_k}{\rho_0}]\rho_q \nonumber \\
    (\chi^{-1})^{2,3} &= \frac{6}{5} \alpha \frac{(3\pi^2)^{2/3}}{\rho_0^{4/3}} ~ \eta^2 ~ [1 - 8\frac{\rho_q}{\rho_0}]\rho_q \nonumber
\end{align}

Similarly, for \textbf{fourth and fifth term}:

\begin{align}
    v_{\theta}^{KGE2,4,5} &= \frac{2}{3} \frac{C_{TF}\alpha\rho^{2/3}}{(1+\alpha s^2)^{3}} [ 3\frac{\nabla^2\rho}{c_s^2\rho^{5/3}}(1-9\alpha s^2) ] \nonumber\\
    & ~~~~~~~~~~\text{Now,}  \nabla^2\rho << \nabla\rho; \nonumber\\
    &= 2\frac{\alpha C_{TF}}{c_s^2\rho}(1-3\alpha s^2)(1-9\alpha s^2) \nabla^2\rho \nonumber\\
    &= -\frac{3}{20}\frac{\alpha}{\rho_0} [1-12\alpha s^2+\mathcal{O}(s^2)]~ q^2\rho_k^2
\end{align}
In the last step we used small perturbation approximations. 
\begin{align}
    \frac{\partial v^{KGE2,4,5}}{\partial\rho_q} &= -\frac{3}{20}\frac{\alpha q^2}{\rho_0} (1 + 3 \frac{\alpha}{(3\pi^2)^2} q^2 \frac{\rho_q^2}{\rho_0^{8/3}} ) \nonumber\\
    (\chi^-1)^{4,5} &= -\frac{3}{20}\frac{\alpha}{\rho_0} 4\eta^2 (3\pi^2\rho_0)^{2/3} [ 1 + 3 \frac{\alpha q^2}{(3\pi^2\rho_0)^{2/3}}  (\frac{\rho_q}{\rho_0})^2]
\end{align}
where in the last step , the final expression of $\chi^{PG}$ is:

\begin{align}
    (\chi^{-1})^{KGE2} &= \frac{\pi^{2}}{k_{F}} - \frac{3}{5}\alpha\eta^2 \frac{\rho_q}{\rho_0} + \frac{18}{5} \frac{\pi^2}{k_F}\alpha \eta^2 (\frac{\rho_q}{\rho_0}) \Big(1 - 8\frac{\rho_q}{\rho_0} \Big) \nonumber\\ &~~~~~~~~~~~~~  - \frac{9}{5} \frac{\pi^2 }{k_F} \eta^2 \Big( 1 + 3\frac{\alpha q^2}{k_{F}^2} (\frac{\rho_q}{\rho_0})^2 \Big)
\end{align}

considering low-s limit, if $\delta\rho=\rho_q\to0$, we get,
\begin{align}
    \chi^{-1} &= \frac{\pi^2}{k_{F}}\{1 - \frac{9}{5}\alpha\eta^2 \}
\end{align}
Putting $\alpha=0$, we can recover TF response function, which is expected.

\twocolumngrid
\bibliography{reference}

@article{KS,
  title = {Self-Consistent Equations Including Exchange and Correlation Effects},
  author = {Kohn, W. and Sham, L. J.},
  journal = {Phys. Rev.},
  volume = {140},
  issue = {4A},
  pages = {A1133--A1138},
  numpages = {0},
  year = {1965},
  month = {Nov},
  publisher = {American Physical Society},
  doi = {10.1103/PhysRev.140.A1133}
}

@article{HK,
  title = {Inhomogeneous Electron Gas},
  author = {Hohenberg, P. and Kohn, W.},
  journal = {Phys. Rev.},
  volume = {136},
  issue = {3B},
  pages = {B864--B871},
  numpages = {0},
  year = {1964},
  month = {Nov},
  publisher = {American Physical Society},
  doi = {10.1103/PhysRev.136.B864},
  url = {https://link.aps.org/doi/10.1103/PhysRev.136.B864}
}

@article{nearsightedness_Kohn_2005,
author = {E. Prodan  and W. Kohn },
title = {Nearsightedness of electronic matter},
journal = {Proceedings of the National Academy of Sciences},
volume = {102},
number = {33},
pages = {11635-11638},
year = {2005},
doi = {10.1073/pnas.0505436102},
URL = {https://www.pnas.org/doi/abs/10.1073/pnas.0505436102}
}

@article{Large-scale-DFT,
    author = {Aarons, Jolyon and Sarwar, Misbah and Thompsett, David and Skylaris, Chris-Kriton},
    title = "{Perspective: Methods for large-scale density functional calculations on metallic systems}",
    journal = {The Journal of Chemical Physics},
    volume = {145},
    number = {22},
    pages = {220901},
    year = {2016},
    month = {12},
    issn = {0021-9606},
    doi = {10.1063/1.4972007},
    url = {https://doi.org/10.1063/1.4972007}
}

@article{review-article-OFDFT-LargeScale,
author = {Mi, Wenhui and Luo, Kai and Trickey, S. B. and Pavanello, Michele},
title = {Orbital-Free Density Functional Theory: An Attractive Electronic Structure Method for Large-Scale First-Principles Simulations},
journal = {Chemical Reviews},
volume = {123},
number = {21},
pages = {12039-12104},
year = {2023},
doi = {10.1021/acs.chemrev.2c00758},
URL = {https://doi.org/10.1021/acs.chemrev.2c00758}
}

@article{Conjointness-conj-PhysRevA.44.768,
  title = {Conjoint gradient correction to the Hartree-Fock kinetic- and exchange-energy density functionals},
  author = {Lee, Hsing and Lee, Chengteh and Parr, Robert G.},
  journal = {Phys. Rev. A},
  volume = {44},
  issue = {1},
  pages = {768--771},
  numpages = {0},
  year = {1991},
  month = {Jul},
  publisher = {American Physical Society},
  doi = {10.1103/PhysRevA.44.768},
  url = {https://link.aps.org/doi/10.1103/PhysRevA.44.768}
}

@article{Hemanadhan_2012,
   title={Response function analysis of excited-state kinetic energy functional constructed by splitting k-space},
   volume={66},
   pages = {57},
   ISSN={1434-6079},
   url={http://dx.doi.org/10.1140/epjd/e2012-20677-4},
   DOI={10.1140/epjd/e2012-20677-4},
   number={2},
   journal={The European Physical Journal D},
   publisher={Springer Science and Business Media LLC},
   author={Hemanadhan, M. and Harbola, M. K.},
   year={2012},
   month=feb 
   }

@book{bookdft_Parr-Yang-1989,
  author    = {R. G. Parr and W. Yang},
  title     = {Density-Functional Theory of Atoms and Molecules},
  publisher = {Oxford University Press},
  address   = {New York},
  year      = {1989},
}

@article{PROFESS:software,
title = {Introducing PROFESS: A new program for orbital-free density functional theory calculations},
journal = {Computer Physics Communications},
volume = {179},
number = {11},
pages = {839-854},
year = {2008},
issn = {0010-4655},
doi = {https://doi.org/10.1016/j.cpc.2008.07.002},
url = {https://www.sciencedirect.com/science/article/pii/S0010465508002476},
author = {Gregory S. Ho and Vincent L. Lignères and Emily A. Carter}
}

@article{DFTpy:Shao_2020,
   title={DFTpy: An efficient and object‐oriented platform for orbital‐free DFT simulations},
   volume={11},
   pages = {e1482},
   ISSN={1759-0884},
   DOI={10.1002/wcms.1482},
   number={1},
   journal={WIREs Computational Molecular Science},
   publisher={Wiley},
   author={Shao, Xuecheng and Jiang, Kaili and Mi, Wenhui and Genova, Alessandro and Pavanello, Michele},
   year={2020}
   }

@article{CALYPSO-1-WANG20122063,
title = {CALYPSO: A method for crystal structure prediction},
journal = {Computer Physics Communications},
volume = {183},
number = {10},
pages = {2063-2070},
year = {2012},
issn = {0010-4655},
doi = {https://doi.org/10.1016/j.cpc.2012.05.008},
url = {https://www.sciencedirect.com/science/article/pii/S0010465512001762},
author = {Yanchao Wang and Jian Lv and Li Zhu and Yanming Ma}
}

@article{CALYPSO-2-PhysRevB.82.094116,
  title = {Crystal structure prediction via particle-swarm optimization},
  author = {Wang, Yanchao and Lv, Jian and Zhu, Li and Ma, Yanming},
  journal = {Phys. Rev. B},
  volume = {82},
  issue = {9},
  pages = {094116},
  numpages = {8},
  year = {2010},
  month = {Sep},
  publisher = {American Physical Society},
  doi = {10.1103/PhysRevB.82.094116},
  url = {https://link.aps.org/doi/10.1103/PhysRevB.82.094116}
}

@article{CALYPSO-3-10.1063/1.4746757,
    author = {Lv, Jian and Wang, Yanchao and Zhu, Li and Ma, Yanming},
    title = "{Particle-swarm structure prediction on clusters}",
    journal = {The Journal of Chemical Physics},
    volume = {137},
    number = {8},
    pages = {084104},
    year = {2012},
    month = {08},
    issn = {0021-9606},
    doi = {10.1063/1.4746757},
    url = {https://doi.org/10.1063/1.4746757}
}

@article{QuantumEspresso-Giannozzi_2009,
doi = {10.1088/0953-8984/21/39/395502},
url = {https://dx.doi.org/10.1088/0953-8984/21/39/395502},
year = {2009},
month = {sep},
publisher = {},
volume = {21},
number = {39},
pages = {395502},
author = {Paolo Giannozzi and Stefano Baroni and Nicola Bonini and Matteo Calandra and Roberto Car and Carlo Cavazzoni and Davide Ceresoli and Guido L Chiarotti and Matteo Cococcioni and Ismaila Dabo and Andrea Dal Corso and Stefano de Gironcoli and Stefano Fabris and Guido Fratesi and Ralph Gebauer and Uwe Gerstmann and Christos Gougoussis and Anton Kokalj and Michele Lazzeri and Layla Martin-Samos and Nicola Marzari and Francesco Mauri and Riccardo Mazzarello and Stefano Paolini and Alfredo Pasquarello and Lorenzo Paulatto and Carlo Sbraccia and Sandro Scandolo and Gabriele Sclauzero and Ari P Seitsonen and Alexander Smogunov and Paolo Umari and Renata M Wentzcovitch},
title = {QUANTUM ESPRESSO: a modular and open-source software project for quantum
simulations of materials},
journal = {Journal of Physics: Condensed Matter}
}

@article{KATO_cusp1,
author = {Kato, Tosio},
title = {On the eigenfunctions of many-particle systems in quantum mechanics},
journal = {Communications on Pure and Applied Mathematics},
volume = {10},
number = {2},
pages = {151-177},
doi = {https://doi.org/10.1002/cpa.3160100201},
url = {https://onlinelibrary.wiley.com/doi/abs/10.1002/cpa.3160100201},
year = {1957}
}

@article{KATO_cusp2,
    author = {Ahlrichs, Reinhart},
    title = "{Asymptotic behavior of atomic bound state wave functions}",
    journal = {Journal of Mathematical Physics},
    volume = {14},
    number = {12},
    pages = {1860-1863},
    year = {1973},
    month = {12},
    issn = {0022-2488},
    doi = {10.1063/1.1666258},
    url = {https://doi.org/10.1063/1.1666258}
}

@article{TF:thomas1927calculation,
  title={The calculation of atomic fields},
  author={Thomas, Llewellyn H},
  journal={Mathematical Proceedings of the Cambridge Philosophical Society},
  volume={23},
  number={5},
  pages={542--548},
  year={1927},
  doi={10.1017/S0305004100011683}
}

@article{JonesAndGunnarsson,
  title = {The density functional formalism, its applications and prospects},
  author = {Jones, R. O. and Gunnarsson, O.},
  journal = {Rev. Mod. Phys.},
  volume = {61},
  issue = {3},
  pages = {689--746},
  numpages = {0},
  year = {1989},
  month = {Jul},
  publisher = {American Physical Society},
  doi = {10.1103/RevModPhys.61.689},
  url = {https://link.aps.org/doi/10.1103/RevModPhys.61.689}
}

@article{Pearson_is_good_10.1063/1.2774974,
    author = {García-Aldea, David and Alvarellos, J. E.},
    title = {Kinetic energy density study of some representative semilocal kinetic energy functionals},
    journal = {The Journal of Chemical Physics},
    volume = {127},
    number = {14},
    pages = {144109},
    year = {2007},
    month = {10},
    issn = {0021-9606},
    doi = {10.1063/1.2774974},
    url = {https://doi.org/10.1063/1.2774974},
}

@article{AugPC_semilocal_KEDF_molecoule_2024,
author = {Wang, Tingwei and Luo, Kai and Lu, Ruifeng},
title = {Semilocal Kinetic Energy Density Functionals on Atoms and Diatoms},
journal = {Journal of Chemical Theory and Computation},
volume = {20},
number = {12},
pages = {5176-5187},
year = {2024},
doi = {10.1021/acs.jctc.4c00532},
URL = {https://doi.org/10.1021/acs.jctc.4c00532}
}

@article{NL-KEDF-PhysRevA.34.2614,
  title = {Explicit estimation of ground-state kinetic energies from electron densities},
  author = {Herring, Conyers},
  journal = {Phys. Rev. A},
  volume = {34},
  issue = {4},
  pages = {2614--2631},
  numpages = {0},
  year = {1986},
  month = {Oct},
  publisher = {American Physical Society},
  doi = {10.1103/PhysRevA.34.2614},
  url = {https://link.aps.org/doi/10.1103/PhysRevA.34.2614}
}

@article{WT_1992,
  title = {Kinetic-energy functional of the electron density},
  author = {Wang, Lin-Wang and Teter, Michael P.},
  journal = {Phys. Rev. B},
  volume = {45},
  issue = {23},
  pages = {13196--13220},
  numpages = {0},
  year = {1992},
  month = {Jun},
  publisher = {American Physical Society},
  doi = {10.1103/PhysRevB.45.13196},
  url = {https://link.aps.org/doi/10.1103/PhysRevB.45.13196}
}

@ARTICLE{WGC_1998,
       author = {{Wang}, Yan Alexander and {Govind}, Niranjan and {Carter}, Emily A.},
        title = "{Erratum: Orbital-free kinetic-energy functionals for the nearly free electron gas [Phys. Rev. B 58, 13 465 (1998)]}",
      journal = {\prb},
         year = 1999,
        month = dec,
       volume = {60},
       number = {24},
        pages = {17162-17163},
          doi = {10.1103/PhysRevB.60.17162},
       adsurl = {https://ui.adsabs.harvard.edu/abs/1999PhRvB..6017162W}
}

@article{WGC_1999,
  title = {Orbital-free kinetic-energy density functionals with a density-dependent kernel},
  author = {Wang, Yan Alexander and Govind, Niranjan and Carter, Emily A.},
  journal = {Phys. Rev. B},
  volume = {60},
  issue = {24},
  pages = {16350--16358},
  numpages = {0},
  year = {1999},
  month = {Dec},
  publisher = {American Physical Society},
  doi = {10.1103/PhysRevB.60.16350},
  url = {https://link.aps.org/doi/10.1103/PhysRevB.60.16350}
}

@article{KGAP_2018,
   title={Nonlocal kinetic energy functional from the jellium-with-gap model: Applications to orbital-free density functional theory},
   volume={97},
   pages = {205137},
   ISSN={2469-9969},
   url={http://dx.doi.org/10.1103/PhysRevB.97.205137},
   DOI={10.1103/physrevb.97.205137},
   number={20},
   journal={Physical Review B},
   publisher={American Physical Society (APS)},
   author={Constantin, Lucian A. and Fabiano, Eduardo and Della Sala, Fabio},
   year={2018},
   month=may }

@article{KGAP_2017,
   title={Jellium-with-gap model applied to semilocal kinetic functionals},
   volume={95},
   pages = {115153},
   ISSN={2469-9969},
   url={http://dx.doi.org/10.1103/PhysRevB.95.115153},
   DOI={10.1103/physrevb.95.115153},
   number={11},
   journal={Physical Review B},
   publisher={American Physical Society (APS)},
   author={Constantin, Lucian A. and Fabiano, Eduardo and Śmiga, Szymon and Della Sala, Fabio},
   year={2017}
   }

@article{Exact-pauli-Levy_PhysRevA.38.625,
  title = {Exact properties of the Pauli potential for the square root of the electron density and the kinetic energy functional},
  author = {Levy, Mel and Hui Ou-Yang},
  journal = {Phys. Rev. A},
  volume = {38},
  issue = {2},
  pages = {625--629},
  numpages = {0},
  year = {1988},
  month = {Jul},
  publisher = {American Physical Society},
  doi = {10.1103/PhysRevA.38.625},
  url = {https://link.aps.org/doi/10.1103/PhysRevA.38.625}
}

@article{Pauli-Density-connection_10.1063/1.4940035,
    author = {Finzel, Kati},
    title = {Local conditions for the Pauli potential in order to yield self-consistent electron densities exhibiting proper atomic shell structure},
    journal = {The Journal of Chemical Physics},
    volume = {144},
    number = {3},
    pages = {034108},
    year = {2016},
    month = {01},
    issn = {0021-9606},
    doi = {10.1063/1.4940035},
    url = {https://doi.org/10.1063/1.4940035}
}

@article{Lieb-Simon-Scaling-1973,
  title = {Thomas-Fermi Theory Revisited},
  author = {Lieb, Elliott H. and Simon, Barry},
  journal = {Phys. Rev. Lett.},
  volume = {31},
  issue = {11},
  pages = {681--683},
  numpages = {0},
  year = {1973},
  month = {Sep},
  publisher = {American Physical Society},
  doi = {10.1103/PhysRevLett.31.681},
  url = {https://link.aps.org/doi/10.1103/PhysRevLett.31.681}
}

@article{PERDEW1992,
title = {Generalized gradient approximation for the fermion kinetic energy as a functional of the density},
journal = {Physics Letters A},
volume = {165},
number = {1},
pages = {79-82},
year = {1992},
issn = {0375-9601},
doi = {https://doi.org/10.1016/0375-9601(92)91058-Y},
url = {https://www.sciencedirect.com/science/article/pii/037596019291058Y},
author = {John P. Perdew}
}

@article{VT84_PhysRevB.88.161108,
  title = {Nonempirical generalized gradient approximation free-energy functional for orbital-free simulations},
  author = {Karasiev, Valentin V. and Chakraborty, Debajit and Shukruto, Olga A. and Trickey, S. B.},
  journal = {Phys. Rev. B},
  volume = {88},
  issue = {16},
  pages = {161108},
  numpages = {5},
  year = {2013},
  month = {Oct},
  publisher = {American Physical Society},
  doi = {10.1103/PhysRevB.88.161108},
  url = {https://link.aps.org/doi/10.1103/PhysRevB.88.161108}
}

@article{PGSL_2018_doi:10.1021/acs.jpclett.8b01926,
author = {Constantin, Lucian
A. and Fabiano, Eduardo and Della Sala, Fabio},
title = {Semilocal Pauli–Gaussian Kinetic Functionals for Orbital-Free Density Functional Theory Calculations of Solids},
journal = {The Journal of Physical Chemistry Letters},
volume = {9},
number = {15},
pages = {4385-4390},
year = {2018},
doi = {10.1021/acs.jpclett.8b01926},
URL = {https://doi.org/10.1021/acs.jpclett.8b01926}
}

@article{PGSL_2019_assessment,
author = {Constantin, Lucian
A. and Fabiano, Eduardo and Della Sala, Fabio},
title = {Performance of Semilocal Kinetic Energy Functionals for Orbital-Free Density Functional Theory},
journal = {Journal of Chemical Theory and Computation},
volume = {15},
number = {5},
pages = {3044-3055},
year = {2019},
doi = {10.1021/acs.jctc.9b00183},
URL = {https://doi.org/10.1021/acs.jctc.9b00183}
}

@article{GSE2_2019_Lucian_PhysRevB.99.155137,
  title = {Semilocal properties of the Pauli kinetic potential},
  author = {Constantin, Lucian A.},
  journal = {Phys. Rev. B},
  volume = {99},
  issue = {15},
  pages = {155137},
  numpages = {12},
  year = {2019},
  month = {Apr},
  publisher = {American Physical Society},
  doi = {10.1103/PhysRevB.99.155137},
  url = {https://link.aps.org/doi/10.1103/PhysRevB.99.155137}
}

@article{LKT-SBTrichey-GGA_PhysRevB.98.041111,
  title = {A simple generalized gradient approximation for the noninteracting kinetic energy density functional},
  author = {Luo, Kai and Karasiev, Valentin V. and Trickey, S. B.},
  journal = {Phys. Rev. B},
  volume = {98},
  issue = {4},
  pages = {041111},
  numpages = {5},
  year = {2018},
  month = {Jul},
  publisher = {American Physical Society},
  doi = {10.1103/PhysRevB.98.041111},
  url = {https://link.aps.org/doi/10.1103/PhysRevB.98.041111}
}

@article{LKTF-SBTrichey-GGA_PhysRevB.101.075116,
  title = {Towards accurate orbital-free simulations: A generalized gradient approximation for the noninteracting free energy density functional},
  author = {Luo, K. and Karasiev, V. V. and Trickey, S. B.},
  journal = {Phys. Rev. B},
  volume = {101},
  issue = {7},
  pages = {075116},
  numpages = {9},
  year = {2020},
  month = {Feb},
  publisher = {American Physical Society},
  doi = {10.1103/PhysRevB.101.075116},
  url = {https://link.aps.org/doi/10.1103/PhysRevB.101.075116}
}

@article{MGP_2018,
    author = {Mi, Wenhui and Genova, Alessandro and Pavanello, Michele},
    title = "{Nonlocal kinetic energy functionals by functional integration}",
    journal = {The Journal of Chemical Physics},
    volume = {148},
    number = {18},
    pages = {184107},
    year = {2018},
    month = {05},
    issn = {0021-9606},
    doi = {10.1063/1.5023926},
    url = {https://doi.org/10.1063/1.5023926},
}

@article{XMW_PhysRevB.100.205132,
  title = {Nonlocal kinetic energy density functional via line integrals and its application to orbital-free density functional theory},
  author = {Xu, Qiang and Wang, Yanchao and Ma, Yanming},
  journal = {Phys. Rev. B},
  volume = {100},
  issue = {20},
  pages = {205132},
  numpages = {9},
  year = {2019},
  month = {Nov},
  publisher = {American Physical Society},
  doi = {10.1103/PhysRevB.100.205132},
  url = {https://link.aps.org/doi/10.1103/PhysRevB.100.205132}
}

@article{LMGP-00,
  title = {Orbital-free density functional theory correctly models quantum dots when asymptotics, nonlocality, and nonhomogeneity are accounted for},
  author = {Mi, Wenhui and Pavanello, Michele},
  journal = {Phys. Rev. B},
  volume = {100},
  issue = {4},
  pages = {041105},
  numpages = {6},
  year = {2019},
  month = {Jul},
  publisher = {American Physical Society},
  doi = {10.1103/PhysRevB.100.041105},
  url = {https://link.aps.org/doi/10.1103/PhysRevB.100.041105}
}

@article{HC,
  title = {Nonlocal orbital-free kinetic energy density functional for semiconductors},
  author = {Huang, Chen and Carter, Emily A.},
  journal = {Phys. Rev. B},
  volume = {81},
  issue = {4},
  pages = {045206},
  numpages = {15},
  year = {2010},
  month = {Jan},
  publisher = {American Physical Society},
  doi = {10.1103/PhysRevB.81.045206},
  url = {https://link.aps.org/doi/10.1103/PhysRevB.81.045206}
}

@article{JGM-10.1063/5.0204957,
    author = {Bhattacharjee, Abhishek and Jana, Subrata and Samal, Prasanjit},
    title = "{First step toward a parameter-free, nonlocal kinetic energy density functional for semiconductors and simple metals}",
    journal = {The Journal of Chemical Physics},
    volume = {160},
    number = {22},
    pages = {224110},
    year = {2024},
    month = {06},
    issn = {0021-9606},
    doi = {10.1063/5.0204957},
    url = {https://doi.org/10.1063/5.0204957}
}

@article{LJGM_JCTC_Bhattacharjee_2026,
    author = {Bhattacharjee, Abhishek and Jana, Subrata and {\'S}miga, Szymon and Samal, Prasanjit},
    title = {Nonlocal Orbital-Free Kinetic Energy Functional from
the Jellium-with-Gap Model for Finite Systems},
    journal = {Journal of Chemical Theory and Computation},
    volume = {22},
    number = {13},
    pages = {6510-6524},
    year = {2026},
    month = {06},
    issn = {1549-9618},
    doi = {10.1021/acs.jctc.6c00460},
    url = {https://doi.org/10.1021/acs.jctc.6c00460}
}

@article{revHC_2021,
  title = {Revised Huang-Carter nonlocal kinetic energy functional for semiconductors and their surfaces},
  author = {Shao, Xuecheng and Mi, Wenhui and Pavanello, Michele},
  journal = {Phys. Rev. B},
  volume = {104},
  issue = {4},
  pages = {045118},
  numpages = {8},
  year = {2021},
  month = {Jul},
  publisher = {American Physical Society},
  doi = {10.1103/PhysRevB.104.045118},
  url = {https://link.aps.org/doi/10.1103/PhysRevB.104.045118}
}

@article{SM_1993,
  title = {Orbital-free kinetic-energy functionals for first-principles molecular dynamics},
  author = {Smargiassi, Enrico and Madden, Paul A.},
  journal = {Phys. Rev. B},
  volume = {49},
  issue = {8},
  pages = {5220--5226},
  numpages = {0},
  year = {1994},
  month = {Feb},
  publisher = {American Physical Society},
  doi = {10.1103/PhysRevB.49.5220},
  url = {https://link.aps.org/doi/10.1103/PhysRevB.49.5220}
}

@Article{BLPS-HC,
author ="Huang, Chen and Carter, Emily A.",
title  ="Transferable local pseudopotentials for magnesium{,} aluminum and silicon",
journal  ="Phys. Chem. Chem. Phys.",
year  ="2008",
volume  ="10",
issue  ="47",
pages  ="7109-7120",
publisher  ="The Royal Society of Chemistry",
doi  ="10.1039/B810407G",
url  ="http://dx.doi.org/10.1039/B810407G"}

@article{Elastic_carter,
  title = {Density-decomposed orbital-free density functional theory for covalently bonded molecules and materials},
  author = {Xia, Junchao and Carter, Emily A.},
  journal = {Phys. Rev. B},
  volume = {86},
  issue = {23},
  pages = {235109},
  numpages = {15},
  year = {2012},
  month = {Dec},
  publisher = {American Physical Society},
  doi = {10.1103/PhysRevB.86.235109},
  url = {https://link.aps.org/doi/10.1103/PhysRevB.86.235109}
}

@article{PhysRevB.85.045126,
  title = {Toward an orbital-free density functional theory of transition metals based on an electron density decomposition},
  author = {Huang, Chen and Carter, Emily A.},
  journal = {Phys. Rev. B},
  volume = {85},
  issue = {4},
  pages = {045126},
  numpages = {9},
  year = {2012},
  month = {Jan},
  publisher = {American Physical Society},
  doi = {10.1103/PhysRevB.85.045126},
  url = {https://link.aps.org/doi/10.1103/PhysRevB.85.045126}
}

@article{10.1063/5.0146167,
    author = {Thapa, Bishal and Jing, Xin and Pask, John E. and Suryanarayana, Phanish and Mazin, Igor I.},
    title = "{Assessing the source of error in the Thomas–Fermi–von Weizsäcker density functional}",
    journal = {The Journal of Chemical Physics},
    volume = {158},
    number = {21},
    pages = {214112},
    year = {2023},
    month = {06},
    issn = {0021-9606},
    doi = {10.1063/5.0146167},
    url = {https://doi.org/10.1063/5.0146167}
}

@article{
BrichMurnaghan,
author = {F. D. Murnaghan },
title = {The Compressibility of Media under Extreme Pressures},
journal = {Proceedings of the National Academy of Sciences},
volume = {30},
number = {9},
pages = {244-247},
year = {1944},
doi = {10.1073/pnas.30.9.244},
URL = {https://www.pnas.org/doi/abs/10.1073/pnas.30.9.244},
}

@article{APBE,
  title = {Semiclassical Neutral Atom as a Reference System in Density Functional Theory},
  author = {Constantin, Lucian A. and Fabiano, E. and Laricchia, S. and Della Sala, F.},
  journal = {Phys. Rev. Lett.},
  volume = {106},
  issue = {18},
  pages = {186406},
  numpages = {4},
  year = {2011},
  month = {May},
  publisher = {American Physical Society},
  doi = {10.1103/PhysRevLett.106.186406},
}

@article{APBEk,
author = {Laricchia, S. and Fabiano, E. and Constantin, L. A. and Della Sala, F.},
title = {Generalized Gradient Approximations of the Noninteracting Kinetic Energy from the Semiclassical Atom Theory: Rationalization of the Accuracy of the Frozen Density Embedding Theory for Nonbonded Interactions},
journal = {Journal of Chemical Theory and Computation},
volume = {7},
number = {8},
pages = {2439-2451},
year = {2011},
month = {07},
doi = {10.1021/ct200382w}
}

@article{Nagy-PauliPot-PhysRevA.84.032506,
  title = {Density and pair-density scaling for deriving the Euler equation in density-functional and pair-density-functional theory},
  author = {Nagy, {\'A}.},
  journal = {Phys. Rev. A},
  volume = {84},
  issue = {3},
  pages = {032506},
  numpages = {5},
  year = {2011},
  month = {Sep},
  publisher = {American Physical Society},
  doi = {10.1103/PhysRevA.84.032506},
  url = {https://link.aps.org/doi/10.1103/PhysRevA.84.032506}
}

@article{Atomic_Pauli_pot_2021,
    author = {Redd, Jeremy J. and Cancio, Antonio C.},
    title = {Analysis of atomic Pauli potentials and their large-Z limit},
    journal = {The Journal of Chemical Physics},
    volume = {155},
    number = {13},
    pages = {134112},
    year = {2021},
    month = {10},
    issn = {0021-9606},
    doi = {10.1063/5.0059283},
    url = {https://doi.org/10.1063/5.0059283}
}

@article{scaling4,
	author = {Nagy, {\'A}.},
    title = {Density scaling and exchange-correlation energy},
    journal = {The Journal of Chemical Physics},
    volume = {123},
    number = {4},
    pages = {044105},
    year = {2005},
    month = {08},
    issn = {0021-9606},
    doi = {10.1063/1.1979473},
    url = {https://doi.org/10.1063/1.1979473},
}

@article{Jones2015,
  title = {Density functional theory: Its origins, rise to prominence, and future},
  author = {Jones, R. O.},
  journal = {Rev. Mod. Phys.},
  volume = {87},
  issue = {3},
  pages = {897--923},
  numpages = {27},
  year = {2015},
  month = {Aug},
  publisher = {American Physical Society},
  doi = {10.1103/RevModPhys.87.897},
}

@article{vasp1,
  title = {Ab initio molecular dynamics for liquid metals},
  author = {Kresse, G. and Hafner, J.},
  journal = {Phys. Rev. B},
  volume = {47},
  issue = {1},
  pages = {558(R)--561(R)},
  numpages = {0},
  year = {1993},
  month = {Jan},
  publisher = {American Physical Society},
  doi = {10.1103/PhysRevB.47.558},
}

@article{PW91,
  title = {Accurate and simple analytic representation of the electron-gas correlation energy},
  author = {Perdew, John P. and Wang, Yue},
  journal = {Phys. Rev. B},
  volume = {45},
  issue = {23},
  pages = {13244--13249},
  numpages = {0},
  year = {1992},
  month = {Jun},
  publisher = {American Physical Society},
  doi = {10.1103/PhysRevB.45.13244},
  url = {https://link.aps.org/doi/10.1103/PhysRevB.45.13244}
}

@article{TD-OFDFT-Vignale-causuality,
  title = {Real-time resolution of the causality paradox of time-dependent density-functional theory},
  author = {Vignale, Giovanni},
  journal = {Phys. Rev. A},
  volume = {77},
  issue = {6},
  pages = {062511},
  numpages = {9},
  year = {2008},
  month = {Jun},
  publisher = {American Physical Society},
  doi = {10.1103/PhysRevA.77.062511},
  url = {https://link.aps.org/doi/10.1103/PhysRevA.77.062511}
}

@article{WDM-1-PhysRevLett.113.155006,
  title = {Fast and Accurate Quantum Molecular Dynamics of Dense Plasmas Across Temperature Regimes},
  author = {Sjostrom, Travis and Daligault, J\'er\^ome},
  journal = {Phys. Rev. Lett.},
  volume = {113},
  issue = {15},
  pages = {155006},
  numpages = {5},
  year = {2014},
  month = {Oct},
  publisher = {American Physical Society},
  doi = {10.1103/PhysRevLett.113.155006},
  url = {https://link.aps.org/doi/10.1103/PhysRevLett.113.155006}
}

@article{WDM-2-PhysRevLett.121.145001,
  title = {Ab Initio Studies on the Stopping Power of Warm Dense Matter with Time-Dependent Orbital-Free Density Functional Theory},
  author = {Ding, Y. H. and White, A. J. and Hu, S. X. and Certik, O. and Collins, L. A.},
  journal = {Phys. Rev. Lett.},
  volume = {121},
  issue = {14},
  pages = {145001},
  numpages = {6},
  year = {2018},
  month = {Oct},
  publisher = {American Physical Society},
  doi = {10.1103/PhysRevLett.121.145001},
  url = {https://link.aps.org/doi/10.1103/PhysRevLett.121.145001}
}

@article{support,
  title = { Supplemental Material for Kinetic energy constructed from exact gradient expansion of second order in uniform gas limit, (including references [16,19,27,28,42])},
  author = {Abhishek Bhattacharjee and Hemanadhan Myneni and Manoj K. Harbola and Prasanjit Samal},
  journal = {Phys. Rev. B},
  volume = {},
  issue = {},
  pages = {},
  numpages = {},
  year = {2026},
  month = {Sep},
  publisher = {American Physical Society},
  doi = {10.1103/grbr-7gwt},
  url = {https://link.aps.org/doi/10.1103/grbr-7gwt}
}

@Article{Review_art_Zhou_2008,
author = {Huajie Chen and Aihui Zhou},
title = {Orbital-Free Density Functional Theory for Molecular Structure Calculations},
journal = {Numerical Mathematics: Theory, Methods and Applications},
year = {2008},
volume = {1},
number = {1},
pages = {1--28},
issn = {2079-7338},
doi = {https://doi.org/2008-NMTMA-6039},
}

@article{Review_article_2024_Ma_Mi_Wang,
author = {Xu, Qiang and Ma, Cheng and Mi, Wenhui and Wang, Yanchao and Ma, Yanming},
title = {Recent advancements and challenges in orbital-free density functional theory},
journal = {WIREs Computational Molecular Science},
volume = {14},
number = {3},
pages = {e1724},
year = {2024},
doi = {10.1002/wcms.1724},
url = {https://wires.onlinelibrary.wiley.com/doi/abs/10.1002/wcms.1724},

}

@article{Na-cluster-Aguado-2005-PhysRevLett.94.233401,
  title = {Anomalous Size Dependence in the Melting Temperatures of Free Sodium Clusters: An Explanation for the Calorimetry Experiments},
  author = {Aguado, Andr\'es and L\'opez, Jos\'e M.},
  journal = {Phys. Rev. Lett.},
  volume = {94},
  issue = {23},
  pages = {233401},
  numpages = {4},
  year = {2005},
  month = {Jun},
  publisher = {American Physical Society},
  doi = {10.1103/PhysRevLett.94.233401},
  url = {https://link.aps.org/doi/10.1103/PhysRevLett.94.233401}
}

@article{Na-melting-Aguado-2006-PhysRevB.74.115403,
  title = {Small sodium clusters that melt gradually: Melting mechanisms in ${\mathrm{Na}}_{30}$},
  author = {Aguado, Andr\'es and L\'opez, Jos\'e M.},
  journal = {Phys. Rev. B},
  volume = {74},
  issue = {11},
  pages = {115403},
  numpages = {11},
  year = {2006},
  month = {Sep},
  publisher = {American Physical Society},
  doi = {10.1103/PhysRevB.74.115403},
  url = {https://link.aps.org/doi/10.1103/PhysRevB.74.115403}
}

@software{DFTpyKGE2,
  author    = {Bhattacharjee, Abhishek and Myneni, Hemanadhan},
  title     = {{DFTpy-KGE2: Kinetic Energy from Exact Second-Order Gradient Expansion}},
  version   = {1.0.0},
  year      = {2026},
  publisher = {Zenodo},
  doi       = {10.5281/zenodo.22667331},
  url       = {https://doi.org/10.5281/zenodo.22667331}
}

\clearpage
\onecolumngrid

\section*{Supplemental Material}
This Supporting Information provides additional details regarding the construction of KGE2, benchmark datasets, density analyses, cluster geometries, and complete numerical results used in the main manuscript.

\begin{align}
    \tau_s &= \tau_{TF}F_{\theta}(s) + \tau_{vW} 
\end{align}
where $s= \frac{|\nabla \rho|}{2k_{F}\rho} = \frac{|\nabla \rho|}{2(3\pi^2)^{1/3}\rho^{4/3}}$,   $\tau_{TF}=\frac{3}{10}(3\pi^2)^{2/3}\rho^{5/3}(\R)$, and $\tau_{vW}(\mathbf{r}) = \frac{|\nabla n(\mathbf{r})|^2}{8\,n(\mathbf{r})}$

\section{LKT Functional \cite{LKT-SBTrichey-GGA_PhysRevB.98.041111}}
The form of \textbf{LKT} functional is :
\begin{align}
    F^{\text{LKT}}_{\theta}(s) &= \frac{1}{\cosh(a s)}
\end{align}
where a=1.3 is fixed.

\section{PG-class of functionals \cite{PGSL_2018_doi:10.1021/acs.jpclett.8b01926}}

The form of \textbf{PG} functional:
\begin{align}
    \label{eq:PG}
    F^{\text{PG}}_{\theta}&= e^{-\mu s^2} 
\end{align}

Here $\mu=40/27=1.48$ gives PGS functional, and $\mu=1.0$ is named as PG1.

%
%

\section{Proposed functional form: KGE2}

Funtional form of \textbf{KGE2} is given by: 
\begin{align}
    \label{eq:KGE2}
    F^{\text{KGE2}}_{\theta}(s) &= \frac{1}{1+\alpha s^2}
\end{align}
where $\alpha=40/27=1.481$ is fixed. 









\begin{figure}[!p]
  \centering
  \includegraphics[width=0.75\textwidth]{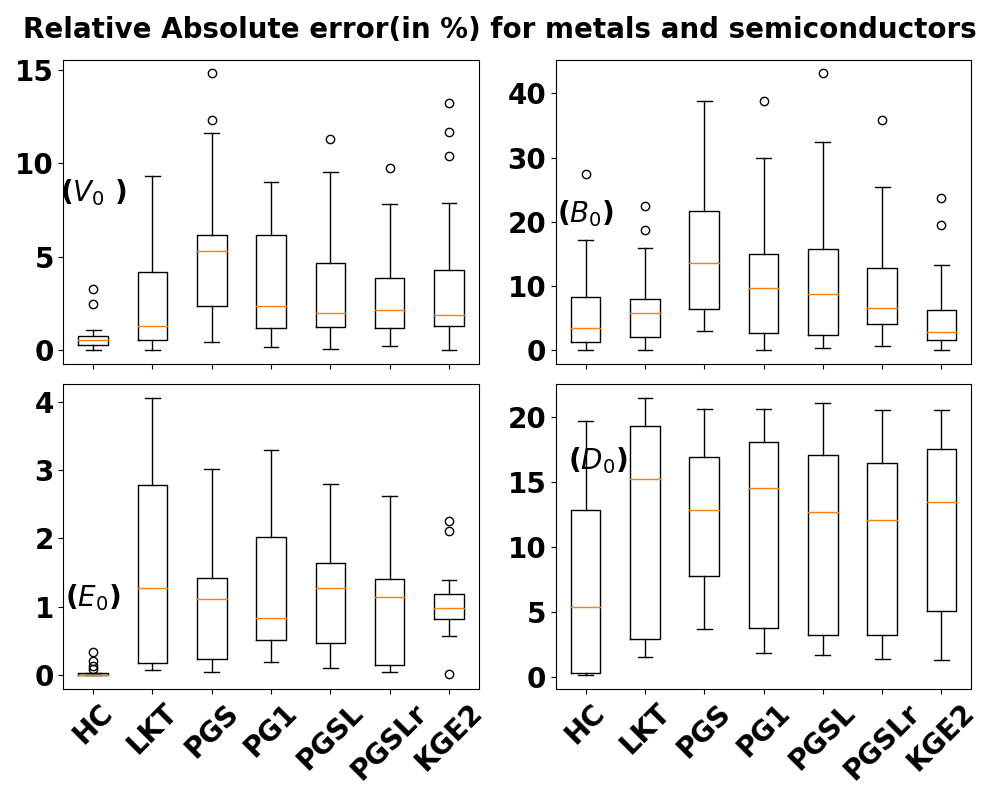}
  \end{figure}
\begin{figure}[!p]
  \centering  
    \includegraphics[scale=0.5]{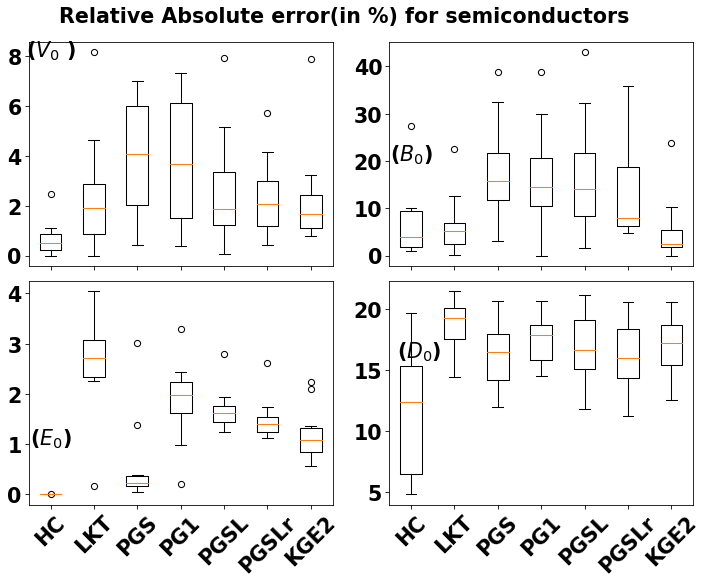}
  \end{figure}
\begin{figure}[!p]
  \centering
   \includegraphics[scale=0.55]{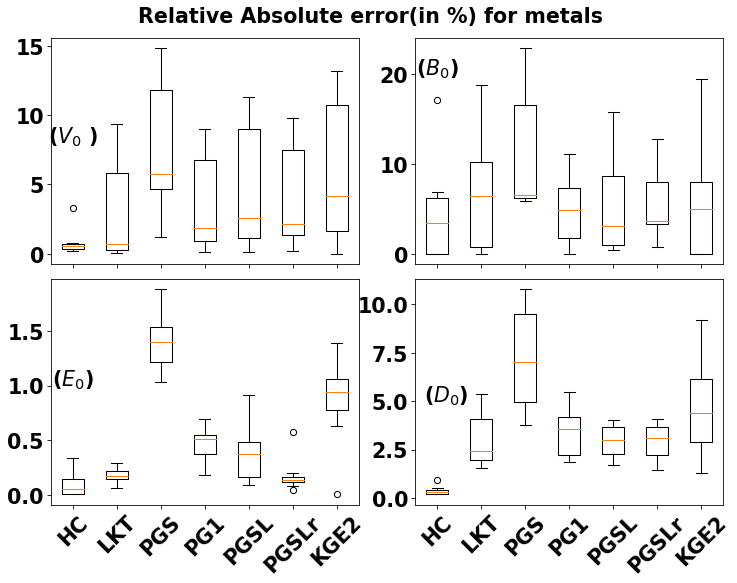}
  \caption{Box plot of relative absolute error (in \%) of equilibrium volume ($V_{0}$), bulk modulus($B_{0}$), energy($E_0$) and density error  ($D_{0}$) for semiconductors and metals. The Box plot used here summarizes the overall distribution of a set of data points in Table.1, Table.2, Table.3 and Table.5 : the bottom and the top of the box are the first (Q1) and third (Q3) quartiles,(25 \% of data points lies below Q1, and another 25 \% above Q3); the band inside the box denotes the median; the circles, if any, denote outliers that lie further than 1.5 $\times$ $ \mid Q3-Q1\mid$ away from the box. The vertical line extends from the minimum to the maximum, except for the outliers. }
  \label{fig:RMARE}
\end{figure}

\begin{figure}
  \centering
   \includegraphics[scale=0.6]{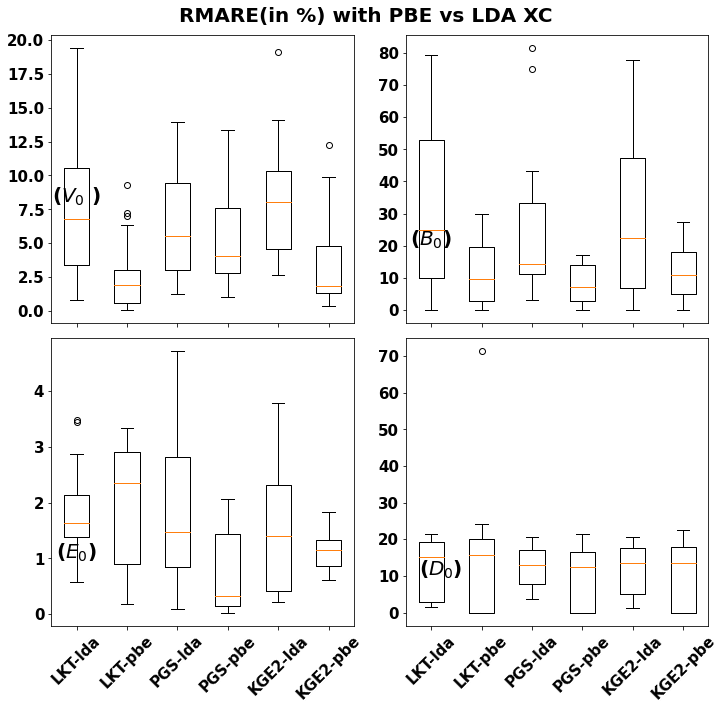}
  \caption{Here we have reproduced Fig.~\ref{fig:RMARE} with LDA and PBE XC.  }
\end{figure}


\clearpage
\begin{table*}
    \caption{ Relative \textbf{Mean} Absolute Error (RmARE) in (\%) of various functionals on Solids.}
    \centering
    \begin{tabular}{cccccccccccc}
    \hline\hline 
    &   & SM & HC &  LKT & PGS & PG1 & KGE2 &  PGSL & PGSLr & PGSL8/81 & KGE28/81 \\
    &  & ($A=0.2$) & ($\lambda=0.011$) & ($a=1.3$) & ($\mu=1.48$) & ($\mu=1$) & ($\alpha=1.48$) & ($\beta=0.25$) & ($\lambda=0.4$) & ($\beta=8/81$) & ($\beta=8/81$) \\
     & & & ($\beta=0.71$) & & & & & & ($\sigma=0.2$)  &  & \\   
  \hline\\
  RmARE (semi+metal)  &  & 6.411 & 1.588 & 6.642 & 6.637 & 6.443 & 5.700 & 6.015 & \textbf{5.221}  & 6.74 & 5.50 \\
  RmARE (semi)  &  & 6.616 & 1.383  & 7.415 & \textbf{4.319} & 6.694 & 4.512  & 6.022 & 5.368  & 4.70 & 5.24 \\
  RmARE (metals) & & 6.013 & 1.986 & 6.499 & 18.703 & 8.030 &   12.142 & 7.492 & \textbf{6.487}  & 14.67 & 9.40 \\
  \hline\hline
    \end{tabular}
    \label{tab:RMARE}
\end{table*}

\begin{table*}
    \caption{ Relative \textbf{Median} Absolute Error (RMARE) in (\%) of various functionals on Solids.}
    \centering
    \begin{tabular}{cccccccccccc}
    \hline\hline 
    &   & SM & HC &  LKT & PGS & PG1 & KGE2 &  PGSL & PGSLr & PGSL8/81 & KGE28/81 \\
    &  & ($A=0.2$) & ($\lambda=0.011$) & ($a=1.3$) & ($\mu=1.48$) & ($\mu=1$) & ($\alpha=1.48$) & ($\beta=0.25$) & ($\lambda=0.4$)  & ($\beta=8/81$) & ($\beta=8/81$) \\
     & & & ($\beta=0.71$) & & & & & & ($\sigma=0.2$)  &  & \\   
  \hline\\
  RMARE (semi+metal)  &  & 6.825 & 1.174 & 6.984 & 7.771 & 7.40 & \textbf{5.822} & 6.984 & 6.370  & 6.89 & 6.51 \\
  RMARE (semi)  &  & 6.738 & 1.261 & 8.723 & \textbf{3.719} & 7.515 & 4.469 & 6.315 & 5.539  & 4.85 & 5.51 \\ 
  RMARE (metals) &  & 5.808 & 2.191 & \textbf{7.241} & 27.075 & 11.992 & 17.976 & 9.749 & 7.840   & 20.75 & 10.48 \\
  \hline\hline
    \end{tabular}
    \label{tab:RMARE2}
\end{table*}

\begin{figure}[h!]
    \centering
    \includegraphics[scale=0.8]{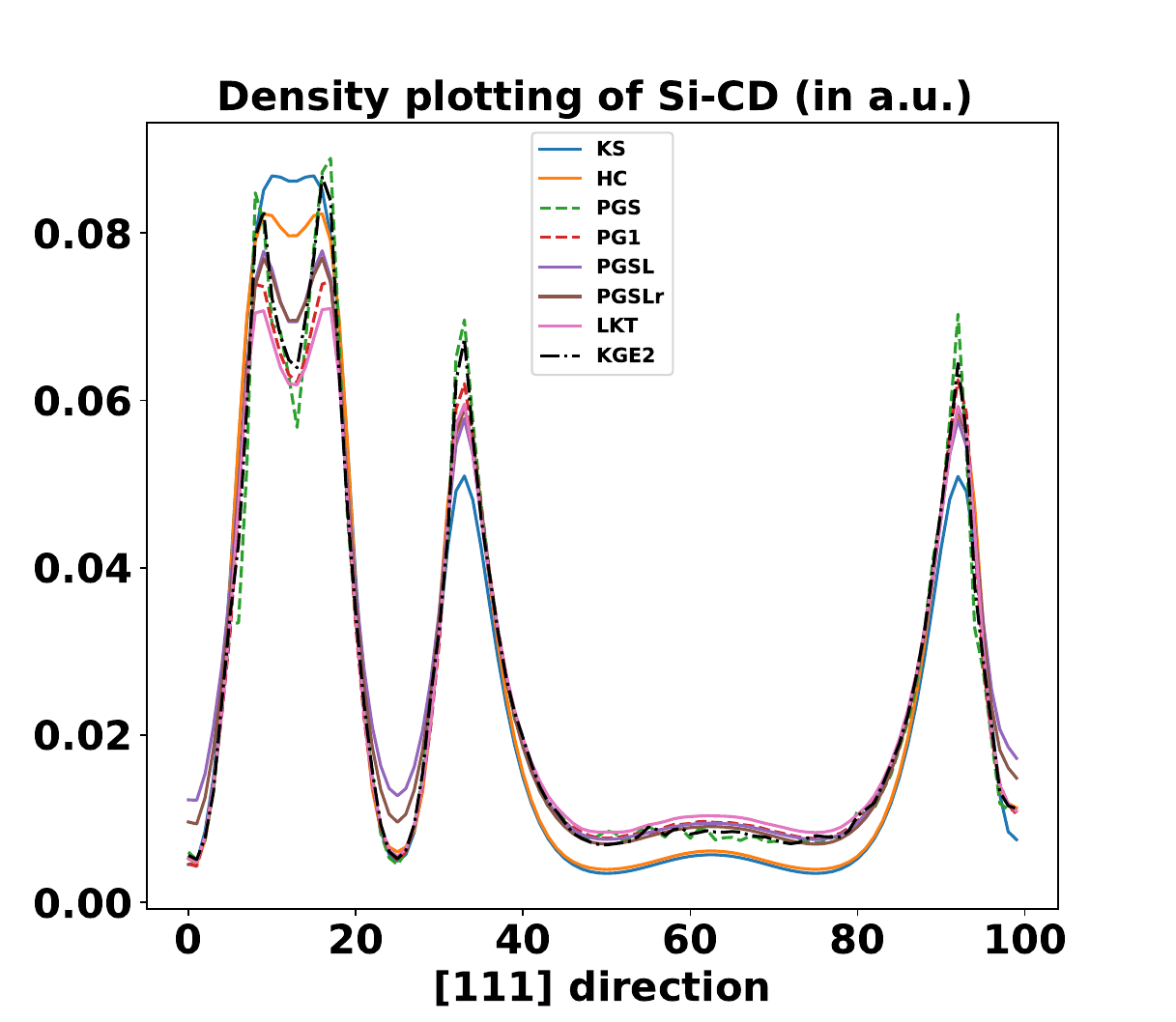}
    \caption{SCF density comparison for Si CD with other functionals. The density is in atomic units and in x-axis we have grid points. }
    \label{fig:enter-label}
\end{figure}

\begin{figure}[h!]
    \centering
    \includegraphics[scale=0.6]{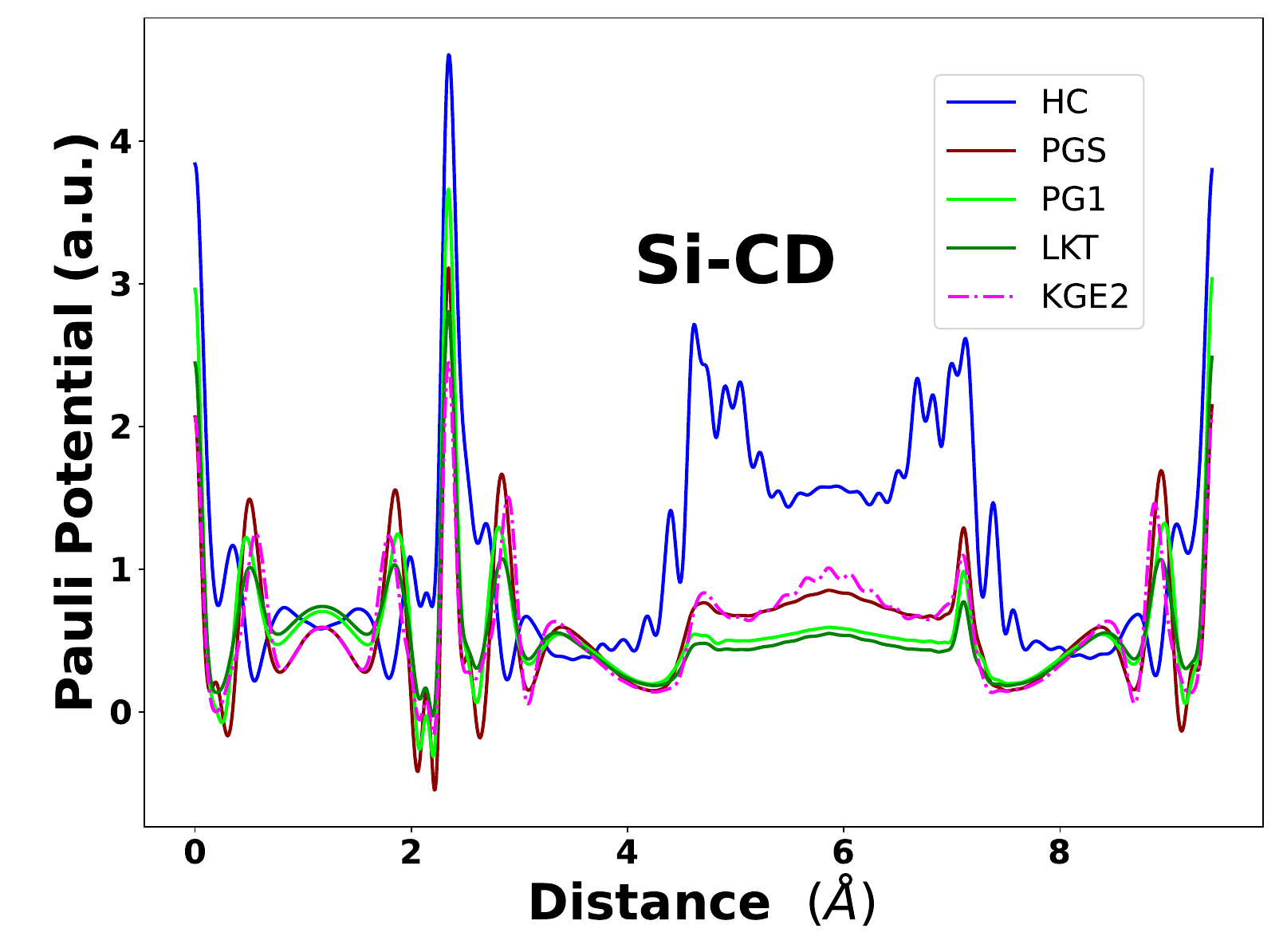}
    \caption{Pauli Potential of various functionals (generated from KS density) plotted and compared. }
    \label{fig:enter-label2}
\end{figure}

\begin{table*}
{\fontsize{8}{8}\selectfont
\caption{\label{tab:semi}
Comparative study of \textbf{KGE2} (Eq.\ref{eq:KGE2}), \textbf{PGS} and \textbf{PG1}  with new generation NL-functionals. Equilibrium volume ($V_0$ in $\AA^3$) per atom, minimum energy ($E_0$ in eV/atom), and bulk modulus ($B_0$ in GPa) for semiconductors. The last column shows the absolute error of density relative to the KS reference density ($D_0$) }
\begin{ruledtabular}
	\begin{tabular}{ccccccc}
 elements & Functional & parameters & $V_0(A^{3}/atom)$    &    $E_0(eV)$  & $B_0(GPa)$ & $D_{0}$ (\%) \\ \hline \hline
 CD-Si & KS \cite{MGP_2018} &  & 19.77 & -109.62 & 98 & \\
   & HC\cite{MGP_2018} & $\lambda=0.001\beta=0.65$ & 19.96 & -109.62 & \textbf{97} & 4.811\\
   & KGAP\cite{KGAP_2018}  & $\alpha=0.65\beta=0.49$ & 19.50 & -108.31 & 77.9 & 7.18 \\
      & LKT &  & 19.77 & -107.09 & 96 & 14.41\\
   & \textcolor{black}{KGE2}  & $\alpha=1.48$ & 19.96 & -109.00 & 88.90 & 12.571 \\
   & \textcolor{black}{PGS}  & $\mu=1.48$ & 20.99 & -109.85 & 60 & 11.95\\
   & \textcolor{black}{PG1} & $\mu=1.0$ & 20.27 & -107.863 &  82 & 14.52 \\
   & \textcolor{black}{PGSLr}  &  & 20.29 & -108.36 & 83 & 11.21 \\
  \hline 
   AlAs & KS \cite{MGP_2018}&  & 21.80 & -117.89 & 80  &\\
   & HC \cite{MGP_2018}& $\lambda=0.012\beta=0.825$ & 22.34 & -117.89 & 76& 11.28\\
   & KGAP\cite{KGAP_2018} & $\alpha=0.85 \beta= 0.48$ & \textbf{21.78} & -114.29 & 67 & 13.63 \\
   & LKT &  & 22.18 & -113.11 & 76 & 17.64\\
   & \textcolor{black}{KGE2}  & $\alpha=1.48$ & 21.99 & -115.24 &  \textbf{78} & 15.84\\
   & \textcolor{black}{PGS}  & $\mu=1.481$ & 23.15 & -116.25 & 66 &15.38\\
   & \textcolor{black}{PG1} & $\mu=1.0$ & 22.80   & -114.017  & 69 & 13.63 \\
   & \textcolor{black}{PGSLr}  & $\alpha=1.48$ & 22.56 & -114.75 & 70 & 14.93  \\
  \hline 
  
  AlP & KS \cite{MGP_2018}&  & 20.31 & -120.09 & 90 &\\
   & HC \cite{MGP_2018}& $\lambda=0.012\beta=0.825$ & \textbf{20.2} & -120.09& \textbf{91} &7.404\\
   & KGAP \cite{KGAP_2018} & $\alpha=0.892\beta=0.478 $ & 19.8 & -117.72 & 78 & 12.69 \\
      & LKT &  & 20.14 & -116.45 & 87 &19.06\\
   & \textcolor{black}{KGE2}  & $\alpha=1.48$ & 20.00 & -118.68 & 87 &16.40\\
   & \textcolor{black}{PGS} & $\mu=1.48$ & 21.11 & -119.71 & 70 &15.18\\
   & \textcolor{black}{PG1} & $\mu=1.0$ & 20.87   & -117.388  & 71 & 17.34 \\
   & \textcolor{black}{PGSLr}  &  &  20.407 & -118.199 & 82 &15.10 \\
  \hline 
  
  AlSb & KS \cite{MGP_2018}&  & 28.3 & -103.303 & 60 &\\
   & HC \cite{MGP_2018}& $\lambda=0.012\beta=0.845$ & 27.99 & -103.29 & 61 &5.974\\
   & KGAP & $\alpha=0.75\beta=0.48 $ & 26.2 & -101.64 & 63 & 11.51 \\
   & LKT &  & 22.18 & -113.11 & 76 &17.535\\
   & \textcolor{black}{KGE2}  & $\alpha=1.48$ & 27.39 & -102.46  & \textbf{59} &15.30\\
   & \textcolor{black}{PGS} & $\mu=1.48$ & 28.42 & -103.19 & 52 &13.75\\
   & \textcolor{black}{PG1} & $\mu=1.0$ & 28.06 & -101.490 & 54 & 15.71 \\
   & \textcolor{black}{PGSLr}  &  & 28.85 & -101.92 & 52 &14.16 \\
  \hline 
    GaAs & KS \cite{MGP_2018}&  & 20.3 & -117.89 & 75 &\\
   & HC \cite{MGP_2018}& $\lambda=0.001\beta=0.783$ & \textbf{20.3} & -117.89 & 81 & 6.15 \\
   & KGAP & $\alpha=0.723\beta=0.487$ & 18.7 & -119.62 & 69 & 9.26 \\
    & LKT &  & 18.64 & -118.08 & 95 &16.18\\
   & \textcolor{black}{KGE2} & $\alpha=1.48$ & 18.70 & -120.37 & 88 &14.36\\
   & \textcolor{black}{PGS} & $\mu=1.48$ & 19.89  & -121.45 & 72 &13.88\\
   & \textcolor{black}{PG1} & $\mu=1.0$ & 19.01 & -119.044 & 92 & 14.6 \\
   & \textcolor{black}{PGSLr}  &  & 21.16 & -116.38 & 71 &12.91 \\
  \hline 
  
  GaP & KS \cite{MGP_2018}&  & 18.8 & -121.52 & 80 &\\
   & HC \cite{MGP_2018}& $\lambda=0.001\beta=0.783$ & \textbf{18.8} & -121.52 & 87 &13.44\\
   & KGAP & $\alpha=0.0.87\beta=0.479$ & 18.27 & -119.30 & 76 & 15.46 \\
   & LKT &  & 18.63 & -118.08 & 90 &19.44\\
   & \textcolor{black}{KGE2} & $\alpha=1.48$ & 18.65 & -120.37 & 85 &17.99\\
   & \textcolor{black}{PGS} & $\mu=1.48$ & 19.84 & -121.46 & 71 &17.56\\
   & \textcolor{black}{PG1} & $\mu=1.0$ & 19.13   & -119.039  & 80 & 18.34 \\
   & \textcolor{black}{PGSLr}  &  & 18.87 & -119.88 & 85 & 16.92 \\
  \hline 
  
  GaSb & KS \cite{MGP_2018}&  & 26.2 & -104.86 & 56 &\\
   & HC \cite{MGP_2018}& $\lambda=0.001\beta=0.783$ & \textbf{26.3} & -104.86 & \textbf{58} &15.71\\
   & KGAP & $\alpha=0.581\beta=0.495$ & 25.5 & -103.70 & 50 & 15.12 \\
   & LKT &  & 25.66 & -102.50 & 62 &19.42\\
   & \textcolor{black}{KGE2} & $\alpha=1.48$ & 25.68 & -104.22 & 61 &18.42\\
   & \textcolor{black}{PGS} & $\mu=1.481$ & 26.76 & -105.02 & 54 &18.105\\
   & \textcolor{black}{PG1} & $\mu=1.0$ & 26.30   & -103.196  & 56 & 18.37 \\
   & \textcolor{black}{PGSLr}  &  &   27.48 & -103.53 & 53 &17.84 \\
   \hline 
  
  InAs & KS \cite{MGP_2018}&  & 24.55 & -114.26 & 65 &\\
   & HC \cite{MGP_2018}& $\lambda=0.001, \beta=0.783$ & 24.7 & -114.26 & 63 &19.70\\
   & KGAP & $\alpha=0.524\beta=0.498$ & 25.9 & -112.95 & 47 & 18.95 \\
    & LKT &  & 25.69 & -110.74 & 61 &21.50\\
   & \textcolor{black}{KGE2} & $\alpha=1.48$ & 25.18 & -112.90 & \textbf{64} &20.55\\
   & \textcolor{black}{PGS} & $\mu=1.481$ & 25.58 & -114.00 & 63 &20.65\\
   & \textcolor{black}{PG1} & $\mu=1.0$ & 26.35   & -111.704  & 55 & 20.66 \\
   & \textcolor{black}{PGSLr}  &  & 25.45 & -112.45 & 61 &20.54 \\
   \hline 
  
  InP & KS \cite{MGP_2018}&  & 23.0 & -117.84 & 73 &\\
   & HC \cite{MGP_2018}& $\lambda=0.001\beta=0.783$ & \textbf{22.9} & -117.84 & 66 &14.85\\
   & KGAP & $\alpha=0.702\beta=0.488$ & 23.29 & -115.80 & 58 & 16.34 \\
   & LKT &  & 23.62 & -113.98 & 67 &20.85\\
   & \textcolor{black}{KGE2} & $\alpha=1.48$ & 23.36 & -116.24 & \textbf{73} &19.124\\
   & \textcolor{black}{PGS} & $\mu=1.48$ & 24.61 & -117.38 & 78 &18.38\\
   & \textcolor{black}{PG1} & $\mu=1.0$ & 24.36   & -114.975  & 59 & 19.52 \\
   & \textcolor{black}{PGSLr}  &  & 24.61 & -116.26 & 72&18.55 \\
   \hline 
  
  InSb & KS \cite{MGP_2018}&  & 31.45 & -101.19 & 50 &\\
   & HC \cite{MGP_2018}& $\lambda=0.001\beta=0.783$ & \textbf{31.4} & -101.19 & 49 &15.51\\
   & KGAP & $\alpha=0.508\beta=0.499$ & 31.67 & -100.26 & 37 & 15.57 \\
   & LKT &  & 31.25 & -98.55 & 49 &20.28\\
   & \textcolor{black}{KGE2} & $\alpha=1.48$ & 30.90 & -100.19 & \textbf{49} &18.75\\
   & \textcolor{black}{PGS} & $\mu=1.48$ & 32.05 & -100.96 & 43 &17.52\\
   & \textcolor{black}{PG1} & $\mu=1.0$  & 31.90   & -99.240   & 44 & 18.84 \\
   & \textcolor{black}{PGSLr}  &  & 31.79 & -99.77 & 46 &18.58 \\
    \\
  \hline 
\end{tabular}
\end{ruledtabular}
}
\end{table*}

\begin{table*}
{\fontsize{8}{7}\selectfont
\caption{\label{tab:metals}
Comparative study of equilibrium volume ($V_0$ in $\AA^3$) per atom , minimum energy ($E_0$ in eV/atom), and bulk modulus ($B_0$ in GPa) of \textbf{KGE2}: Eq.\ref{eq:KGE2}, \textbf{PGS} and \textbf{PG1} with new generation NL-functionals on metals. Note that for metals, KGAP becomes SM functional \cite{KGAP_2018} and $\lambda=0$ for HC with same argument mentioned in \cite{HC}.   }
\begin{ruledtabular}
	\begin{tabular}{ccccccc}
 elements & Functional & parameters & $V_0(A^{3}/atom)$    &    $E_0(eV)$  & $B_0(GPa)$ &$D_0$ (\%) \\ \hline \hline
   Al-BCC & KS \cite{XMW_PhysRevB.100.205132} &  & 16.05 & -57.97 & 75 & \\
   & HC  & $\beta=0.65$ & 15.96 & -57.98 & 78 &0.36 \\
   & SM/KGAP  & $\alpha=\beta=0.5$ & 15.13& -57.98 & 78 & 0.63\\
      & LKT &  & 16.87 & -58.02 & 87 &2.27 \\
   & \textcolor{black}{KGE2} & $\alpha=1.48$ & 17.72 & -58.58 & 72 &7.39 \\
   &\textcolor{black}{PGS}& $\mu=1.48$ & 18.03 & -58.81 & 64 & 9.37 \\
   & \textcolor{black}{PG1} & $\mu=1.0$ & 17.03 & -58.22 & 80 & 3.71\\
   &\textcolor{black}{PGSLr}&  & 17.23 & -58.00 & 72 & 3.09 \\
  \hline 
  
     Al-FCC & KS \cite{XMW_PhysRevB.100.205132} &  & 15.60 & -57.94 & 83 & \\
   & HC  & $\beta=0.65$ & 15.96 & -57.86 & 72 & 0.22 \\
   & SM/KGAP & $\alpha=\beta=0.5$ & 14.74 & -58.10 & 86 & 1.31 \\
      & LKT &  & 16.82 & -58.04 & 88 &3.96 \\
   &\textcolor{black}{KGE2}& $\alpha=1.48$  & 17.66 & -58.60 & 72 &1.31 \\
   &\textcolor{black}{PGS}& $\mu=1.48$  & 17.91 & -58.83 & 64 & 9.98 \\
   & \textcolor{black}{PG1} & $\mu=1.0$  & 16.97 & -58.242 & 81 & 4.77\\
   &\textcolor{black}{PGSLr}&  & 17.12 & -58.03 & 72 & 4.06 \\
  \hline 
  Al-SC & KS \cite{XMW_PhysRevB.100.205132} &  & 18.8 &  -57.58 & 64 & \\
   & HC & $\lambda=0.001\beta=0.65$ & 19.57 & -57.61 & 49  & 0.92 \\
   & SM/KGAP  & $\alpha=\beta=0.5$  & 18.46 & -57.767 & 60 & 3.44 \\
      & LKT &  & 18.79 & -57.46 & 76 & 5.37 \\
   &\textcolor{black}{KGE2}& $\alpha=1.48$ & 19.39 & -58.21 & 67 &3.86 \\
   &\textcolor{black}{PGS}& $\mu=1.48$ & 19.78 & -58.50 & 60 & 5.25 \\
   & \textcolor{black}{PG1} & $\mu=1.0$  & 19.01 & -57.729 & 70 & 4.02 \\
   &\textcolor{black}{PGSLr}&  & 19.31 & -57.66 & 65 & 3.59 \\
  \hline 
   Mg-BCC & KS \cite{XMW_PhysRevB.100.205132} &  & 21.34 & -24.70 & 36 & \\
   & HC & $\beta=0.65$ & 21.49 & -24.62 & 35 & 0.18 \\
   & SM/KGAP  & $\alpha=\beta=0.5$ & 19.78 & -24.74 & 37 & 2.04 \\
      & LKT &  & 23.33 & -24.66 & 33 &4.35 \\
   &\textcolor{black}{KGE2}& $\alpha=1.48$ & 23.83 & -24.86 & 29 &9.19 \\
   &\textcolor{black}{PGS}& $\mu=1.48$ & 23.82 & -24.96 & 28 & 10.80 \\
   & \textcolor{black}{PG1} & $\mu=1.0$ & 23.26 & -24.750 & 32 & 5.48 \\
   &\textcolor{black}{PGSLr}&  & 23.01 & -24.562 & 31 & 3.97 \\
  \hline 
    Mg-FCC & KS \cite{XMW_PhysRevB.100.205132} &  & 21.34 & -24.687 & 37 & \\
   & HC & $\beta=0.65$ & 21.41 & -24.64 & 35 &0.541 \\
   & SM/KGAP  & $\alpha=\beta=0.5$ & 19.88 & -24.76 & \textbf{38} & 1.54 \\   
      & LKT &  & 23.19 & -24.67 & 33 &4.22 \\
   &\textcolor{black}{KGE2}& $\alpha=1.48$ & 23.66 & -24.86 & 29 & 8.59 \\
   &\textcolor{black}{PGS}& $\mu=1.48$ & 23.63 & -24.96 & 27 & 10.12 \\
   & \textcolor{black}{PG1} & $\mu=1.0$ & 23.09 & -24.755 & 32 & 5.22 \\
   &\textcolor{black}{PGSLr}&  & 22.74 & -24.57 & 31 &3.14 \\
  \hline 
   Mg-SC & KS \cite{XMW_PhysRevB.100.205132} &  & 24.89 & -24.303 & 29 & \\
   & HC & $\beta=0.65$  &  25.08 & -24.307 & 27 & 0.30 \\
   & SM/KGAP  & $\alpha=\beta=0.5$  & 24.22 & -24.35 & 27 & 1.22 \\
      & LKT &  & 24.96 & -24.36 & 31 &2.05 \\
   &\textcolor{black}{KGE2}& $\alpha=1.48$ & 25.20 & -24.64 & \textbf{29} &4.90 \\
   &\textcolor{black}{PGS}& $\mu=1.48$ & 25.19 & -24.76 & 27 & 7.27 \\
   & \textcolor{black}{PG1} & $\mu=1.0$ & 24.93 & -24.471 & 30 & 2.33 \\
   &\textcolor{black}{PGSLr}&  & 25.03 & -24.32 & 29 & 2.43 \\
  \hline 
   Li-BCC & KS \cite{XMW_PhysRevB.100.205132} &  & 18.82 & -7.59 & 16 & \\
   & HC & $\beta=0.65$ & 18.77 & -7.59 & 16 &0.52 \\
   & SM/KGAP  & $\alpha=\beta=0.5$ & 18.52 & -7.61 &  17 & 0.5 \\
      & LKT &  & 18.79 & -7.61 & 16 & 1.59 \\
   &\textcolor{black}{KGE2}& $\alpha=1.48$ & 18.10 & -7.66 & 17 &2.76 \\
   &\textcolor{black}{PGS}& $\mu=1.48$ & 17.80 & -7.69 & 17 & 3.75 \\
   & \textcolor{black}{PG1} & $\mu=1.0$ & 18.51 & -7.637 & 17 & 1.92 \\
   &\textcolor{black}{PGSLr}&  & 18.52 & -7.58 & 17 & 1.44 \\
  \hline 
    Li-FCC & KS \cite{XMW_PhysRevB.100.205132}&  & 18.78 & -7.60 & 17 &  \\
   & HC & $\beta=0.65$ & \textbf{18.71} & -7.59 & 16 & 0.33 \\
   & SM/KGAP  & $\alpha=\beta=0.5$ & 18.43 & -7.61 & 17 & 0.51 \\
      & LKT &  & 18.68 & -7.61 & 17 & 1.53 \\
   &\textcolor{black}{KGE2}& $\alpha=1.48$ & 17.94 & -7.67 & \textbf{17} &2.90 \\
   &\textcolor{black}{PGS}& $\mu=1.48$ & 17.65 & -7.69 & 18 & 4.01 \\
   & \textcolor{black}{PG1} & $\mu=1.0$ & 18.40 & -7.640 & 17 & 1.86 \\
   &\textcolor{black}{PGSLr}&  & 18.44 & -7.58 & 17 &1.65 \\
  \hline 
  Li-SC & KS \cite{XMW_PhysRevB.100.205132} &  & 19.41 & -7.458 & 17 & \\
   & HC & $\beta=0.65$  & \textbf{ 19.45} & -7.456 & \textbf{17} &0.21 \\
   & SM/KGAP  & $\alpha=\beta=0.5$  & 19.19 & -7.47 & 17 & 0.84 \\
      & LKT &  & 19.57 & -7.48 & 17 &2.60 \\
   &\textcolor{black}{KGE2}& $\alpha=1.48$ & 19.06 & -7.53 & 18 & 5.74 \\
   &\textcolor{black}{PGS}& $\mu=1.48$ & 18.82 & -7.56 & 18 & 6.82 \\
   & \textcolor{black}{PG1} & $\mu=1.0$ & 19.35 & -7.504 & 17 & 3.45 \\
   &\textcolor{black}{PGSLr}&  & 19.51 & -7.44 & 16 &3.05 \\
\end{tabular}
\end{ruledtabular}
}
\end{table*}

\begin{table*}[h!]
\caption{Elastic constants \(C_{11}\), \(C_{12}\), and \(C_{44}\) for CD Si and other III-V semiconductors calculated by KSDFT and OFDFT with the HC, MGP, WT, KGAP and the current semi-local KEDFs. For HC and KGAP we have used optimal choices of the parameters mentioned in \cite{HC,KGAP_2018} respectively. (* denotes unstable structure.)}
\begin{tabular}{ cccccc }  
  \hline
  elements & Functional & parameters & \(C_{11}(GPa)\) & \(C_{12}(GPa)\) & \(C_{44}(GPa)\) \\
  \hline\hline
   Si-CD  & KS (\cite{Elastic_carter}) &  & 162 (163) & 65 (66)  & 76 (102) \\
    & WGC * \cite{Elastic_carter} &  & 3 & 78 & -150 \\
    & WGCD * \cite{Elastic_carter} &  & 89 & 103 & 52 \\
    & HC & \(\lambda=0.01\) & 101 & 92 & 81 \\
    & KGAP * & Eg=1.17 & 46 & 98 & 77 \\
    & TFvW *&  & 23 & 142 & 77 \\
    & KGE2 *& \(\alpha=1.481\) & 58 & 122 & 35 \\
    & PG1 *& \(\alpha=1.481\) & 2 & 301 & -142  \\
  \hline 
    AlAs & KS &  & 118 & 61 & 55 \\
    & HC * & \(\lambda=0.01\) & 45 & 85 & -19 \\
    & KGAP *& Eg=2.23 & 15 & 67 & 50 \\
    & TFvW  *&  & 15 & 107 & 54 \\
    & KGE2 * & \(\alpha=1.481\) & 58 & 131 & 107 \\
    & PG1 & \(\alpha=1.481\) & 797 & 15 & 602  \\
  \hline 
  AlP & KS &  & 133 & 69 & 61 \\
    & HC *& \(\lambda=0.01\) & 80 & 93 & 107 \\
    & KGAP * & Eg=2.5 & 23 & 63 & 50 \\
    & TFvW *&  & 23 & 126 & 79 \\
    & KGE2 *& \(\alpha=1.481\) & 311 & -37 & -55 \\
    & PG1 * & \(\alpha=1.481\) & -297 & 429 & 236  \\
    \hline
  AlSb & KS &  & 84 & 43 & 38 \\
    & HC *& \(\lambda=0.01\) & 50 & 55 & 44 \\
    & KGAP * & Eg=1.69 & 16 &  41 & 35 \\
    & TFvW *&  & 25 & 85 & 44 \\
    & KGE2 *& \(\alpha=1.481\) & 153 & 19 & 41 \\
    & PG1 *& \(\alpha=1.481\) & -59 & 104 & 31  \\
    \hline
    GaAs & KS &  & 116 & 54 & 56 \\
    & HC & \(\lambda=0.01\) & 34 & 60 & 21 \\
    & KGAP * & Eg=1.52 & 4 & 49 & 34 \\
    & TFvW *&  & -10 & 86 & 38 \\
    & KGE2 *& \(\alpha=1.481\) & 32 & 61 & 42 \\
    & PG1 *& \(\alpha=1.481\) & -342 & 341 & -592 \\
    \hline
    GaP & KS &  & 141 & 67 & 67 \\
    & HC *& \(\lambda=0.01\) & 12 & 89 & -39 \\
    & KGAP *& Eg=2.35 & 13 & 89  & 59 \\
    & TFvW *&  & 0 & 115 & 66 \\
    & KGE2 & \(\alpha=1.481\) & 96 & 31 & 19 \\
    & PG1 & \(\alpha=1.481\) & 349 & -4 & 54 \\
    \hline
    GaSb & KS &  & 86 & 40 & 41 \\
    & HC & \(\lambda=0.01\) & 51 & 49 & 43 \\
    & KGAP *& Eg=0.81 & 12 &  45 & 36 \\
    & TFvW  *&  & 19 & 72 & 49 \\
    & KGE2 & \(\alpha=1.481\) & 20 & 63 & 26 \\
    & PG1 *& \(\alpha=1.481\) & 248 & 58 & -44 \\
    \hline
    InAs & KS &  & 84 & 49 & 36 \\
    & HC *& \(\lambda=0.01\) & 34 & 44 & 22 \\
    & KGAP *& Eg=0.42 & 14 & 44  & 34 \\
    & TFvW *&  & 17 & 62 & 65 \\
    & KGE2 & \(\alpha=1.481\) & 29 & 61 & 79 \\
    & PG1 *& \(\alpha=1.481\) & -176 & 131 & 120 \\
    \hline
    InP & KS &  & 99 & 58 & 42 \\
    & HC *& \(\lambda=0.01\) & 31 & 94 & 37 \\
    & KGAP *& Eg=1.42 & 14 & 63 & 41 \\
    & TFvW *&  & -2 & 85 & 40 \\
    & KGE2 & \(\alpha=1.481\) & 132 & 60 & 92 \\
    & PG1 & \(\alpha=1.481\) & 1719 & 86 & 1051 \\
    \hline
    InSb & KS &  & 65 & 39 & 26 \\
    & HC *& \(\lambda=0.01\) & 33 & 39 & 25 \\
    & KGAP *& Eg=0.24 & 14 & 33 & 31 \\
    & TFvW *&  & 8 & 54 & 26 \\
    & KGE2 *& \(\alpha=1.481\) & -34 & 64 & 7 \\
    & PG1 & \(\alpha=1.481\) & 220 & -38 & 33 \\
    \hline
\end{tabular}
\label{tab:Elastic}
\end{table*}

\begin{table*}[h!]
    {\fontsize{10}{15}\selectfont
    \caption{Benchmarking of energy values of various cluster with various KEDFs }
    \centering
	    \begin{tabular}{cccccccccc}
    \hline
  elements  & KS   & TFvW  & WT   & LWT  & LMGP  & PGS  & KGE2  & LKT & HC \\
  \hline
  Li8       & -5.69   & -4.72   & -7.55   & -5.17   & -5.80   & -5.37   & -5.25   & -5.11   & -5.784 \\
  Li30      & -6.62   & -6.08   & -7.31   & -6.31   & -6.712  & -6.50   & -6.43   & -6.32   & -6.666 \\
  Mg8       & -23.15  & -20.88  & -26.21  & -21.99  & -23.40  & -22.54  & -22.21  & -21.83  & -23.330 \\
  Mg30      & -23.21  & -21.50  & -25.62  & -22.35  & -23.47  & -23.03  & -22.76  & -22.35  & -23.345 \\
  Mg50      & -22.27  & -20.71  & -24.31  & -21.50  & -22.57  & -22.30  & -22.02  & -21.57  &  -22.428 \\
  Al8    & -52.078 & -51.0619 & - & - & -53.31923 & -51.72821 & -50.9339 & -51.72821 & -53.280 \\  
  Si8       & -105.8  & -96.58  & -139.29 & -100.37 & -107.46 & -103.96 & -102.73 & -100.69  & -106.394 \\
  Si30      & -107.366 & -100.84 & -119.57 & -103.72 & -108.13 & -106.79 & -105.78 & -104.00  & -107.723 \\
  Ga4As4    & -113.78 & -106.45 & -152.51 & -110.56 & -117.69 & -114.14 & -112.81 & -110.97  & -116.905 \\
  Ga10As10  & -82.66  & -76.08  & -119.63 & -80.75  & -88.20  & -85.84  & -84.29  & -81.21   & -87.885 \\
  Ga4P4     & -118.99 & -108.088 & -152.020 & -119.296 & -119.295 & -116.436 & -115.006 & -116.437  & -119.183 \\
  Ga4Sb4    & -93.51  & -85.659  & -126.394 & -95.701  & -95.701  & -93.258  & -89.795  & -93.258   & -95.645 \\
  In4As4    & -107.228 & -98.321 & -137.364 & -109.550 & -109.550 & -106.582 & -102.852 & -106.582  & -109.386 \\
  In4P4     & -113.06 & -101.46 & -147.27 & -106.04 & -114.17 & -110.24 & -108.75 & -106.29  & -116.241 \\
  In4Sb4    & -92.855 & -83.985 & -119.772 & -93.637 & -93.636 & -91.239 & -87.931 & -91.239  & -93.531 \\
  Al4P4     & -116.771 & -105.879 & -155.765 & -116.139 & -116.138 & -113.875 & -112.546 & -113.875  & -116.241 \\
  Al4Sb4    & -92.777 & -84.33 & -124.36 & -88.17 & -94.91 & -91.57 & -90.41 & -88.26  & -93.945 \\
  \hline
    \end{tabular}
    \label{tab:my_label}
    }
\end{table*}

\begin{table*}[h!]
    {\fontsize{10}{15}\selectfont
    \caption{ Absolute error of $D_{0}$ for various functionals on cluster systems.}
    \centering
	    \begin{tabular}{ ccccccccccc }
    \hline
    & elements & TFvW & WT & LWT & LMGP & HC & KGE2 & PGS & LKT \\
      \hline
      & Li8 & 15.61  & 30.36  & 9.74  & 10.41  & 10.92 & 13.70 & 13.75  & 13.16  \\
      & Li30 & 12.49  & 13.75  & 6.32  & 4.41  & 4.51 & 9.63 & 9.65  & 9.62  \\
      & Mg8  & 20.95 & 10.70 & 9.17 & 6.57 & 6.38 & 13.86 & 12.85  & 14.83  \\
      & Mg30 & 13.83 & 10.60 & 6.85 & 4.04 & 3.89 & 10.29 & 10.74  & 10.07  \\
      & Mg50 & 12.49 & 10.60 & 6.60 & 3.86 & 3.49 & 8.55 & 9.05 & 8.46  \\
      & Al8 & 15.82  & 21.92  & 8.84  & 6.97  & 6.65 & 11.79 & 11.60  & 12.06  \\      
      & Si8   & 17.79 & 2.58 & 9.63 & 6.37  & 6.21 & 12.99 & 12.43 & 13.68  \\
      & Si30  & 15.11 & 2.19 & 8.04 & 3.82 & 3.53 & 10.21 & 10.27  & 10.63  \\
      & Ga4As4 & 20.19 & 1.13 & 9.56 & 7.38 & 7.43 & 15.20 & 14.85 & 15.36  \\
      & Ga10As10 & 16.41 & 1.89 & 8.91 & 5.70 & 5.16 & 10.92 & 10.52 & 11.67   \\
      & Ga4P4 & 21.69  & 16.56  & 11.65  & 7.51  & 7.61 & 16.02 & 16.02  & 16.55  \\
      & Ga4Sb4 & 19.90  & 9.04  & 9.91  & 7.38  & 7.61 & 14.35 & 14.35  & 15.12  \\
      & In4As4  & 20.00 & - & 10.16 & 7.21 & 7.438 & 14.85 & 14.66 & 15.28 \\
      & In4P4 & 22.38 & 5.60 & 12.78 & 7.99 & 8.54 & 16.43 & 15.10 & 17.31  \\
      & In4Sb4 & 20.75  & 7.34  & 10.82  & 8.20  & 8.38 & 14.85 & 14.85  & 15.76  \\
      & Al4P4 & 19.81  & 27.62  & 10.07  & 5.90  & 6.10 & 14.65 & 13.66  & 15.43  \\
      & Al4Sb4 & 22.03 & 10.30 & 12.75 & 9.41 & 9.71 & 16.00 & 14.62 & 17.05  \\
  \hline
    \end{tabular}
    \label{tab:D0_Clusters_semi}
    }
\end{table*}

\begin{figure*}
    \centering
    \includegraphics[scale=0.3]{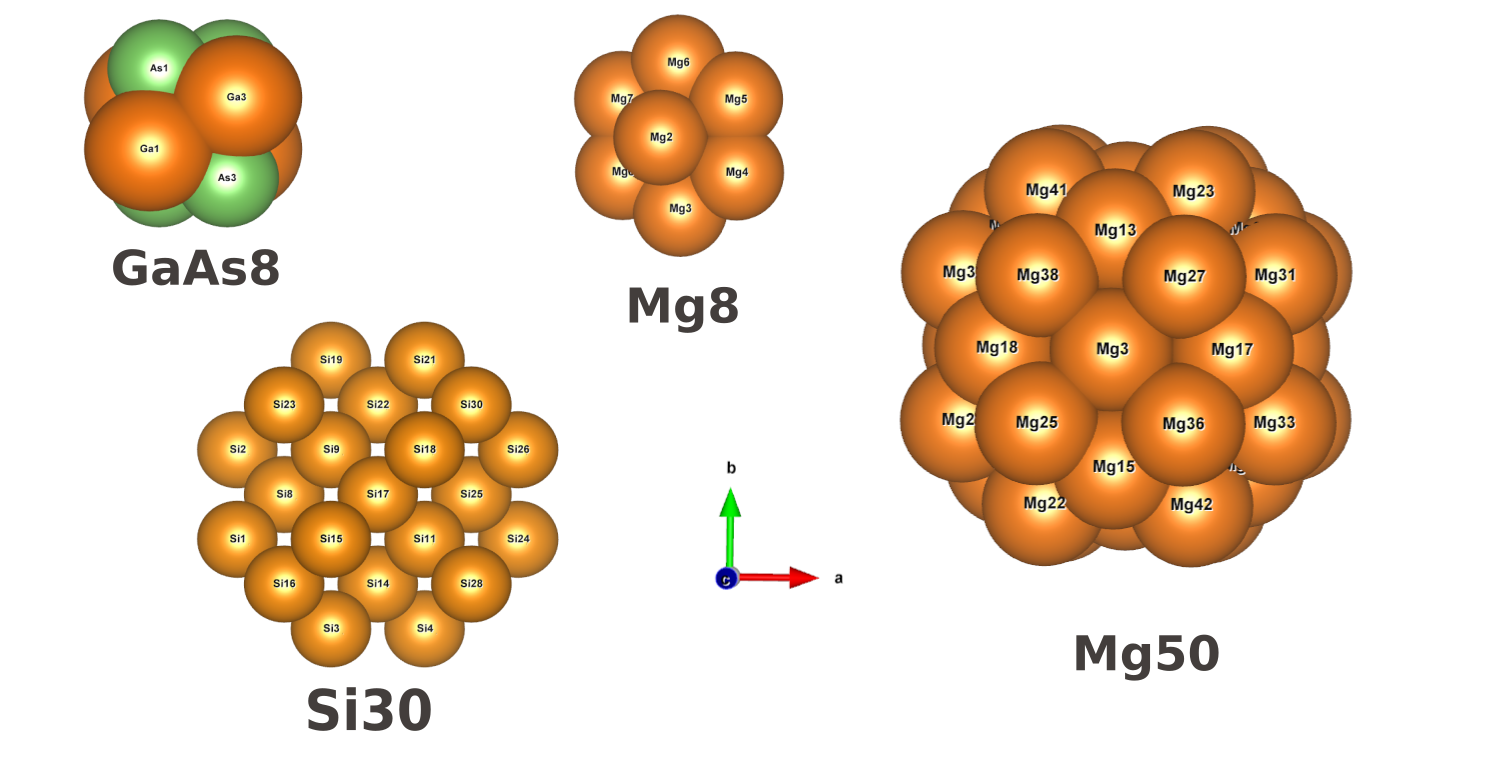}
    \caption{ Few cluster structures used for calculation.}
    \label{fig:cluster_structure}
\end{figure*}

\twocolumngrid

\end{document}